\documentclass[preprint]{aastex}
\usepackage{psfig}

\newcommand{\myemail}{ettoref@oapa.astropa.unipa.it}
\slugcomment{To appear in ApJ}
\shorttitle{The Orion Nebula Cluster. Paper I}
\shortauthors{Flaccomio et al.}

\begin{document}
\title{{\em Chandra} X-ray Observation of the Orion Nebula Cluster. I Detection, Identification and Determination of X-ray Luminosities}
\author{E. Flaccomio\altaffilmark{1}}
\affil{Dipartimento di Scienze Fisiche ed Astronomiche --  Universit\`a di 
Palermo}
\email{\myemail}

\author{F. Damiani, G. Micela,  S. Sciortino}
\affil{INAF - Osservatorio Astronomico di Palermo G.S. Vaiana, Palazzo
dei Normanni, I-90134 Palermo, Italy}

\and

\author{F. R. Harnden Jr., S. S. Murray, S. J. Wolk} 
\affil{Harvard-Smithsonian Center for Astrophysics, 60 Garden Street, 
Cambridge, MA 02138} 

\altaffiltext{1}{Now at INAF - Osservatorio Astronomico di Palermo G.S. Vaiana}

\begin{abstract} 

In this first of two companion papers on the Orion Nebula Cluster (ONC), we
present our analysis of a 63 Ksec {\em Chandra} HRC-I observation that yielded
742 X-ray detections within the $30^{\prime}\times 30^{\prime}$ field of view.
To facilitate our interpretation of the X-ray image, here we collect a
multi-wavelength catalog of nearly 2900 known objects in the region by
combining 17 different catalogs from the recent literature.  We define two
reference groups: an {\em infrared sample}, containing all objects detected in the
$K$ band, and an {\em optical sample} comprising low extinction, well characterized
ONC members. We show for both samples that field object contamination is
generally low.

Our X-ray sources are primarily low mass ONC members. The detection rate for
optical sample stars increases monotonically with stellar mass from zero at the
brown dwarf limit to $\sim 100\%$ for the most massive stars but shows a
pronounced dip between 2 and 10 solar masses.

We determine $L_X$ and $L_X/L_{bol}$ for all stars in our {\em optical sample}
and utilize this information in our companion paper to study correlations
between X-ray activity and other stellar parameters.

\end{abstract} 


\section{Introduction}

The Orion region is one of the most frequently observed areas of the sky. It
comprises several molecular clouds and stellar associations, among which the
Orion Nebula Cluster (ONC - also referred to as the Trapezium region), is of
particular interest.  With an age of about 1 million years, more than two
thousand young stellar objects with masses between 0.08 and 50$M_\odot$, an
estimated central density of $\sim 2 \times 10^4$ stars per cubic parsec, and
at a distance of $\sim 470$pc, it is the largest and densest concentration of
young pre-main sequence (PMS) stars in our region of the Galaxy. Only part of
the ONC members -- those lying on  the near side of the Orion Molecular Cloud
from which the cluster is forming -- are optically visible; more than half are
so deeply embedded in the cloud that they are observable only at infrared (IR)
or X-ray wavelengths where the cloud becomes more transparent. For more
complete descriptions of the ONC structure, dynamics, and stellar content, see
\citet{hil97,hil98b,hil00,car00,luh00,ode01}, and references therein. Among
others, Hillenbrand and collaborators \citep{hil97,hil98a,hil00} have
extensively studied both the optically visible and the embedded ONC stellar
content. 
Several studies \citep{sta99,her00,her01,reb01} have measured
rotational periods  via photometric modulation of stellar spots. Spectacular
HST images \citep[see e.g.,][]{bal00} have allowed the direct observation of
many circumstellar disks and {\em proplyds} (photoionization structures due to
the evaporation of disks) that seem to surround a large fraction of the PMS
stars in the vicinity of the central bright star $\theta^1$ Orionis C.

The ONC was first detected in the X-rays by the {\em Uhuru} satellite \citep{gia72}. Following imaging observations performed with the {\em Einstein} and ROSAT observatories \citep{ku79,gag95a} indicated that: {\em a}) ONC members are powerful X-ray emitters, with typical $L_X/L_{bol}$ close to the saturation value, $10^{-3}$, observed for fast rotators on the main sequence; {\em b}) no relation seems to hold between X-ray activity and stellar rotation; {\em c}) stars with circumstellar accretion disks may have lower levels of X-ray activity. These works were however hindered by the lack, at that time, of complete optical information on the ONC population and by the low sensitivity and spatial resolution of the X-ray instrumentation, especially important to resolve the dense central region and to unambiguously identify X-ray sources with optical counterparts. Both of these limitations have been recently lifted: {\em Chandra} observations \citep{sch01,gar00} have revealed about 1000 X-ray point sources, for the most part associated with low mass stars down to the substellar mass limit.

In this work we describe the analysis of a {\em Chandra} High Resolution Camera
\citep[HRC, ][]{mur00} observation of the ONC and correlate our data with
relevant data from the literature. In a companion work (\citealt{fla02};
hereafter Paper~II), we investigate statistical correlations between X-ray
activity and other stellar characteristics (such as rotational period, mass,
age and disk accretion indicators) and search for insights into the physical
mechanisms that drive activity in PMS stars. An investigation of the
X-ray variability characteristics of ONC members based on these data is in
preparation.

The structure of this paper is as follows: we first introduce our {\em Chandra}
observation and discuss its reduction and X-ray source detection (\S
\ref{sect:obs}). We then describe the preparation of an extensive catalog of
known X-ray/optical/IR/radio objects in the region of our HRC observation,
collecting from the literature useful observable properties and defining two
reference object samples (\S \ref{sect:cat}).
In the same section we also report the results of our observation in the BN/KL region.  
In \S \ref{sect:optIRprop} we
concentrate on the optical/IR properties of our X-ray sources, and in \S
\ref{sect:Xprop} and a supporting Appendix, we determine the X-ray activity indicators $L_X$ and
$L_X/L_{bol}$, along with their uncertainties. Our findings are summarized in
\S \ref{sect:disc}, setting the stage for Paper~II.

\section{Observation and Data Reduction \label{sect:obs}}

The HRC on board the {\em Chandra} X-ray Observatory \citep{wei02}
observed the ONC for 63.2 ksec on 2000 February 4.  The pointing ($R.A._{2000}=
5^h:35^m:17^s$, $DEC._{2000}=-5^\circ:23^{\prime}:16^{\prime\prime}$) was chosen to
place the Trapezium region and the bright O star $\theta^1$ Orionis C in the
center of the field of view (FOV). A good fraction of the ONC region was
included in the $30^{\prime} \times 30^{\prime}$ HRC FOV.

Our {\em Chandra} observations offer these significant improvements over
previous {\em Einstein} and ROSAT observations: unprecedented spatial
resolution ($\sim 0.5^{\prime\prime}$ in the field center), important in such a
crowded field (see \S \ref{sect:hrd_id}); sensitivity $\sim 20$ times deeper
than previous X-ray data (see \S \ref{sect:Xprop}); and continuous temporal
coverage for variability studies.

\subsection{Data Filtering \label{sect:filtering}}

Our HRC analysis begins by filtering the event list to lower background
levels\footnote{Comprising $\sim 80\%$ of the total count rate, background dominates the data.}
and increase the source signal-to-noise ratio (SNR). We considered filtering
out particular time intervals and/or ranges of pulse height amplitude ($PHA$)
dominated by instrumental and spacecraft environment effects. We examined light
curves of the entire observation but found that our particular data
set did {\em not} suffer from variable solar activity contamination commonly reported for
{\em Chandra} data. Hence no time filtering was applied. On the other hand, $PHA$
distributions of all events do show a sharp instrumental peak for $PHA$ values
$\lesssim 35$, and as in our previous HRC analyses \citep{frh01}, we excluded
this peak by eliminating events with $PHA < 35$.
Given the $PHA$ response of the HRC to
X-rays and stellar spectra (cf. \S \ref{sect:conv_fact}),
these counts are almost exclusively background: total events were reduced by
about $15\%$ (from $\sim 2.1$ to $\sim 1.8 \times 10^6$), while source counts
declined only $\sim 1\%$.

\begin{figure}
\centerline{\psfig{figure=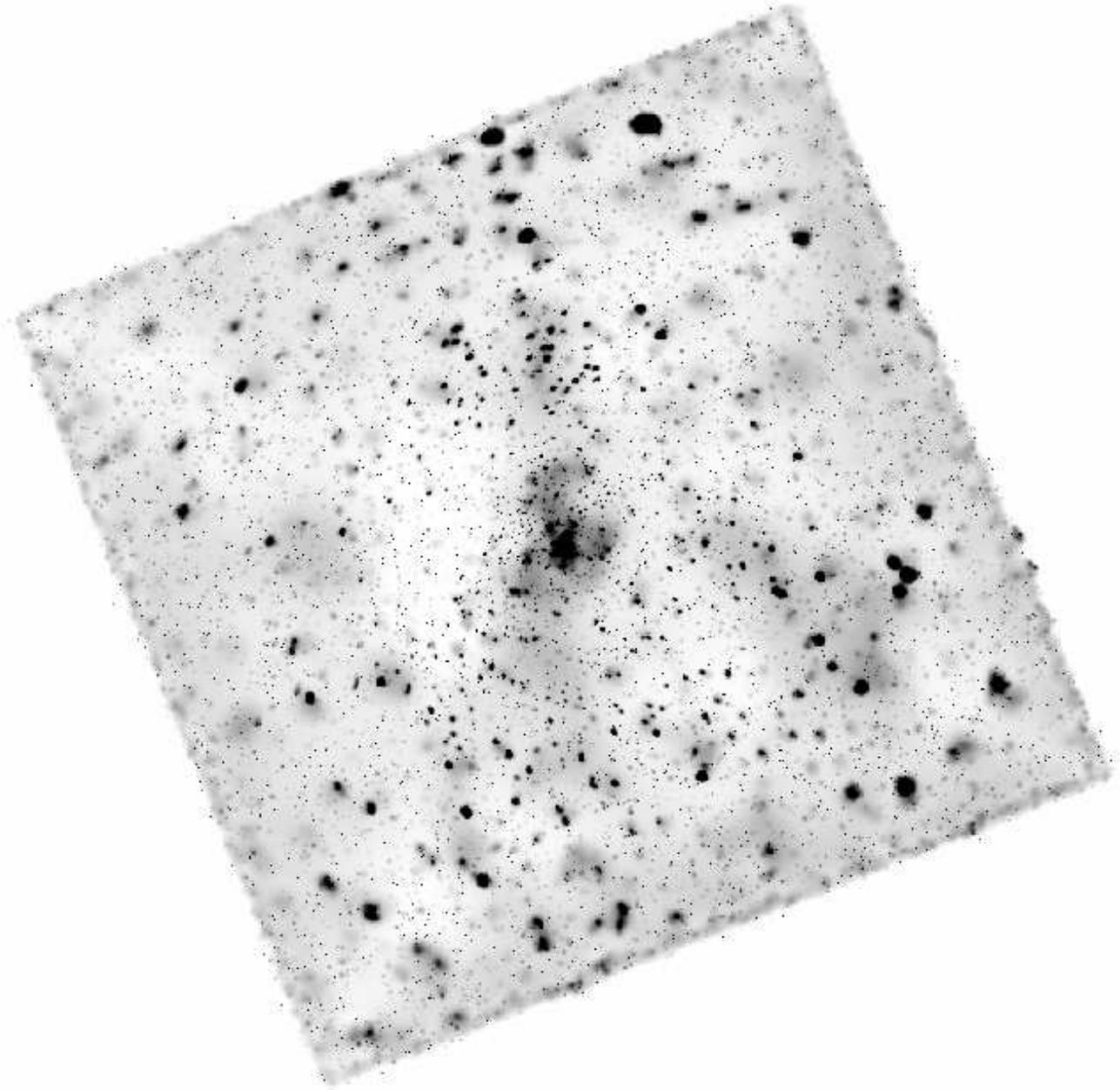,width=18cm}}
\caption{HRC-I image of the ONC processed with {\sc csmooth}. Linear gray scale
indicates photon arrival rate vs. position on the detector. In this
low-resolution presentation black indicates scale saturation due to intense
point-like sources (or unresolved groups of sources). Note that the PSF
increases significantly toward image edges. \label{fig:HRC}}
\end{figure}

Figure \ref{fig:HRC} shows the filtered HRC image, smoothed for presentation
purposes with the {\sc csmooth} task in the {\em Chandra} Interactive Analysis
of Observations ({\sc CIAO}) package, available at $\rm
http://asc.harvard.edu/ciao/$.

\subsection{Source detection \label{sect:detections}}

We have analyzed the filtered event list with the Palermo Wavelet Transform
detection code, {\sc pwdetect} \citep{dam97,dam02}.\footnote{See also {
http://www.astropa.unipa.it/progetti\_ricerca/PWDetect}} In synthesis, {\sc
pwdetect} analyzes an image at a variety of spatial scales, allowing detection
of both point-like and moderately extended sources. The most important parameter
required by the code is the detection threshold which determines, along with
the image background level, the number of expected spurious detections. The
relationship between detection threshold and expected number of spurious sources
was determined by running the {\sc pwdetect} code on 400 simulated background
images with roughly the same number of photons as in the background of our
observation ($\sim 1.46 \times 10^6$, once the source contribution is
removed). 

We choose to accept detections with $SNR > 4.74$, corresponding to 10 expected
spurious detections throughout the FOV.  We accept 10 spurious sources, instead
of the more customary one, for three reasons: i) 10 is still a small fraction
of the total number of detected sources, ii) by lowering the threshold from
$5.25 \sigma$ (corresponding to 1 spurious source) to 4.74, we gain $\sim 100$
{\em good} detections (after a ``hand filtering'' of the detection list -- see
below), of which only 9 are expected to be spurious, and iii) as shown in \S
\ref{sect:hrd_id}, {\em Chandra's} superb spatial resolution ensures that the
$\sim 10$ spurious sources are extremely unlikely to be identified with other
wavelength counterparts.

Visual inspection of the initial lists of $\sim 830$ detections produced by
{\sc pwdetect} showed that although the vast majority of detections appear real
and most of the obvious sources in the image are detected, two ``glitches'' of
the detection algorithm produce undesired effects. First, about 90 sources are detected in the vicinity of very bright stars,
notably on the wings of the point spread function (PSF) of the central source
($\theta$ $^1$Ori C). These are most likely spurious detections due to large
intensity fluctuations on the wings of bright sources. In such regions the code
is unable to estimate correctly the highly spatially variable background level.
Second, in a few cases ($\sim 20$) a pair of sources clearly visible and
distinguishable by eye in the X-ray image is detected as both a point source
corresponding to one of the two and an extended source encompassing the pair.
We dealt with the first problem (to be addressed in a future version of {\sc
pwdetect}) by visually examining the list of detections and excluding the
spurious sources in question. Specifically, we deleted all sources lying closer
than 6.8" to $\theta^1$ Orionis C, thereby also losing any real sources
occurring in that area. The second problem occurs because the extended
detection has a higher SNR than the point-like detection. For such cases, we
eliminated this problem by running a special version of {\sc pwdetect} that
limits its search to point-like sources.

{\sc pwdetect} estimates positions and count rates of detected sources in the
assumption that the PSF is Gaussian. Since the actual HRC PSF is not Gaussian,
we investigated the effect of this simplifying assumption on our estimates. We
computed the wavelet transform of calibration PSFs (in the {\em Chandra}
Calibration Database; see {\em asc.harvard.edu/caldb}) at several detector
positions and for a representative energy of $\sim 1.5$ keV. In order to better
match the actual shape of bright sources in our HRC image, these calibration
PSFs were first convolved with a bi-dimensional Gaussian with
$\sigma=0.3^{\prime\prime}$, possibly reflecting the effect of imperfect aspect
reconstruction affecting the real source PSF. Corrections respect to the
Gaussian PSF approximation were parametrized as a function of off-axis angle
and, for count rates, range from 16\% at the field center to 0\% at $\gtrsim
10^\prime$ off-axis. Having verified that this result agrees well with the
(background-subtracted) extracted counts of bright isolated sources in the
limit of large extraction radii, we applied the correction to our source
count rates. Positional errors (up to $0.9^{\prime\prime}$ for off-axis angles
of $\sim 14^\prime$) are instead always much smaller respect to random
uncertainties and were therefore neglected.


The final lists of detections, comprising 742 X-ray sources, is presented in
Table \ref{tab:detections} where we give sky position and its statistical
uncertainty, detection SNR, number of source counts and uncertainty, and
effective exposure time at the source positions. This latter quantity,
describing the spatially varying sensitivity of the HRC plus {\em Chandra}
mirror system, is derived from an exposure map calculated with {\sc CIAO} for
an  incident energy of 2.0 keV, i.e., the approximate temperature of our
sources (see \S \ref{sect:conv_fact}). This choice of temperature is not
critical for our purposes, however, because the normalized effective area at
any given point on the detector depends only marginally on energy: for $0.5 <
kT < 3.5$, the values of effective exposure times at any given location on the
detector vary at most at the $\sim$4\% level.

\begin{deluxetable}{ccccrrrc}
\tabletypesize{\scriptsize}
\tablewidth{0pt}
\tablecaption{X-ray sources detected in the Orion Nebula Cluster  \label{tab:detections}}
\tablehead{
\colhead{$N_X$}&   \colhead{$Ra (2000)$}&   \colhead{$Dec (2000)$}&   \colhead{$\delta_{Ra,Dec}$}&   \colhead{$SNR$}&   \colhead{$Counts$}&   \colhead{$\delta_{cnts}$}&   \colhead{$Exp. T.$} \\
\colhead{}&   \colhead{[$~h$ $~m$ $~s$]}&   \colhead{[$^\circ$ \ $^{\prime}$ \ $^{\prime\prime}$]}&   \colhead{[$^{\prime\prime}$]}&   \colhead{}&   \colhead{}&   \colhead{}&   \colhead{[ks]} \\
\colhead{(1)}&
\colhead{(2)}&
\colhead{(3)}&
\colhead{(4)}&
\colhead{(5)}&
\colhead{(6)}&
\colhead{(7)}&
\colhead{(8)}}
\startdata
  1&   5  34 14.46&  -5  28 16.62&    3.38&   36.27&   1277.75&     67.97&     49.06\\
  2&   5  34 18.23&  -5  33 28.53&    6.91&    6.65&    167.54&     41.78&     47.64\\
  3&   5  34 20.70&  -5  30 46.81&    5.90&    9.21&    259.90&     46.23&     49.44\\
  4&   5  34 20.74&  -5  23 25.85&    8.03&    4.76&    111.66&     33.69&     51.22\\
  5&   5  34 24.96&  -5  22  6.32&    3.17&   42.89&   1095.36&     46.16&     52.24\\
  6&   5  34 27.23&  -5  24 22.23&    2.02&   63.91&   1866.58&     64.05&     52.90\\
  7&   5  34 27.69&  -5  31 55.39&    3.17&   42.05&   1536.32&     72.59&     50.23\\
  8&   5  34 28.53&  -5  24 58.95&    2.20&   50.53&   1225.93&     49.62&     53.21\\
  9&   5  34 29.09&  -5  14 32.51&    4.46&   15.73&    434.32&    225.67&     49.69\\
 10&   5  34 29.30&  -5  23 56.09&    2.30&   45.86&   1024.30&     45.09&     53.52\\
\enddata

\tablecomments{The 742 entries of the full table are available in electronic form at:
http://www.astropa.unipa.it/$\sim$ettoref/PaperI\_tables.html}
\end{deluxetable}

In addition to measuring count rates for detected X-ray sources, we also used
{\sc pwdetect}
to compute upper limits to the photon flux of all non-detected
objects in the catalog described in the next section. This calculation was
performed consistently using the same SNR threshold as for detections 
and applying the correction, described above, due to the non-Gaussian PSF.

\section{The general catalog of known objects \label{sect:cat}}

We assembled an extensive catalog of known X-ray/Optical/IR and radio objects 
that fell within the HRC FOV. In addition to our list of HRC sources and the {\em
Chandra} source lists of \citet{gar00} and \citet{sch01}, we considered 14
catalogs from recent publications, producing a database of nearly 2900 distinct
objects reported in at least one of the studies considered. A full list of
references is given in the first column of Table \ref{tab:master} along with a
concise classification of the work (col. 2) and the referenced table number(s)
from the original work (col. 3).

\subsection{Catalogs cross-identifications}

Cross identifications were based on positional coincidence.  First of all we
registered the coordinates of each catalog to a standard reference list
\citep[that of][]{jon88} using the mean relative displacement of well
identified  objects. In the case of the \citet{gar00} list of X-ray sources, we
also had to apply a rotation of 0.21$^\circ$ around the HRC field center in
order to bring the coordinates into agreement with those of the other lists. We
then performed the final identifications using the following tolerances:  a) For our
list of HRC X-ray sources, the positional uncertainties computed by {\sc
pwdetect} as a function of source statistics and off-axis angle. These range
from $0.036^{\prime\prime}$ for the bright central source ($\theta^1$  Ori C)
to $\sim 8^{\prime\prime}$ for weak sources close to the detector border. b)
For the {\em Chandra} sources of \citet{gar00}, uncertainties were assumed to
follow the same off-axis trend as measured in our data.  We performed a fourth
order polynomial fit of HRC position errors vs. off-axis angle and then
conservatively used twice these values for the \citet{gar00} source off-axis
angles (to allow for additional counting statistics effects). c) For all
other catalogs, we used $\sim 2.5$ times the mean dispersion of the registered position
offsets with respect to the reference list \citep{jon88}. In all cases, this
identification radius was $\lesssim 1.0^{\prime\prime}$. 

We then examined the resulting list of identifications by eye and modified 
some identifications according to more subjective criteria. For example, 9
X-ray sources detected at large off-axis ($\gtrsim 8^{\prime}$), and therefore
with relatively large PSFs, were associated with one or two optical objects
that fell outside the error circle. In these cases these optical objects lay
well within the PSF and thus potentially contributed the detected photons.  In
some of these cases the source centroid lay between two objects unresolved (by {\sc
pwdetect}) so neither was within the formal error circle. In
all such cases we added the missing identifications manually.

\begin{deluxetable}{llrcccrr}
\tabletypesize{\scriptsize}
\tablecaption{List of catalogs in the ONC area. \label{tab:master}}
\tablewidth{0pt}
\tablehead{
\colhead{Reference} & \colhead{Type of study}& \colhead{Table}& \colhead{Num\tablenotemark{a}}&   
\colhead{$R_{95}$\tablenotemark{b}}& 
\colhead{$N_R$\tablenotemark{c}}&   \colhead{$N_{HRC}$\tablenotemark{d}}&  \colhead{Data\tablenotemark{e}}     
}
\startdata
  This work   			& X-ray (HRC-I)	     	&   1&  742		    & 2.2\tablenotemark{\dag}& 553&   - & HRC-I Cnt. rate  	    \\
 \citet{sch01}			& X-ray	(ACIS-S)     	&   1&  111		    & 0.4		  &  63&  92 & ACIS-S Cnt. rate 	    \\
 \citet{gar00}			& X-ray (ACIS-I)     	&   1&  971\tablenotemark{\ddag}& 1.4		  & 559& 579 & -			    \\
 \citet{bal00}			& Optical (HST)     	& 1,2&   48		    & 0.5		  &  17&  14 & -			    \\
 \citet{hil97}			& Optical Phot./Spec.	& 1,3& 1516		    & 0.4		  &1055& 627 & $V^1I^2$ $A_V^2$ $T_{eff}$ $L_{bol}$   \\
 \citet{hil98a}\tablenotemark{f}& IR Photometry	     	&   1& 1516		    &	-		  &   -& 627 & $J^2H^3K^2L^2$ $EW_{CaII}$ $\Delta_{I-K}$\\
 \citet{hil00}\tablenotemark{g}	& IR Photometry	     	&   1&  776		    & 0.4		  & 242& 282 & $H^1K^1$ 			    \\
 \citet{luc00}			& IR Photometry     		&   1&  557		    & 0.7 		  & 102& 116 & $I^1J^1H^2$			    \\
 \citet{luc01}			& IR Spectroscopy     		& 1,3&   23		    &   - 		  &    &   2 & $A_V^1$ $Mass$			    \\
 \citet{lad00}			& IR Phot. Protostars  	&   1&   78		    & 0.5		  &  11&  26 & $L^1$			    \\
 2Mass	      			& IR Photometry	     	&   -& 1596		    & 0.6		  & 728& 416 & $J^4H^5K^4$		    \\
 \citet{car01}			& IR Phot. / Variability  & 4,7&  555		    & 0.5		  & 379& 241 & $J^3H^4K^3$ $P_{rot}^4$      \\
 \citet{sta99}			& $P_{rot}$	     	&   1&  124		    & 0.5		  & 115&  84 & $P_{rot}^3$		    \\
 \citet{her00}\tablenotemark{h} & $P_{rot}$	     	&   2&  132		    &  -		  &    & 109 & $P_{rot}^2$		    \\
 \citet{her01}\tablenotemark{k} & $P_{rot}$	     	&   -&  296		    &  -		  &    & 168 & $P_{rot}^1$		    \\
 \citet{jon88}			& Proper Motion	     	&   3& 1034		    &  -		  &    & 519 & $P_{memb}$		    \\
 \citet{fel93}			& Radio		     	& 1,5&   49		    & 0.9		  &  18&  25 & -			    \\
 Reid (private comm.)			& Radio 	     	&   -&  100		    & 0.4		  &  32&  43 & -			    \\
\enddata

\tablenotetext{a}{Number of sources in the HRC field of view.}
\tablenotetext{b}{95\% quantiles, in arcseconds, of the object distances from the \citet{jon88} counterparts}
\tablenotetext{c}{Number of objects used to derive $R_{95}$, i.e., identified with \citet{jon88}}
\tablenotetext{d}{Number of objects identified with an HRC X-ray source.}
\tablenotetext{e}{Data collected from each catalog; superscripts give precedence for acceptance in our catalog.}
\tablenotetext{f}{Same catalog as in \citet{hil97}}
\tablenotetext{g}{Covers only the central $5'\times5'$ of the HRC FOV}
\tablenotetext{h}{Subset of the catalog of \citet{jon88}}
\tablenotetext{k}{Data provided by W. Herbst. The original list contained 403 stars; we used data only for those in common with \citet{jon88}}
\tablenotetext{\dag}{In the central $5'\times5'$ of the FOV, where the
{\em Chandra} PSF is narrower, $R_{95} \sim 0.65"$, based on 148 objects.}
\tablenotetext{\ddag}{The original lists contains 973 sources. We deleted two that had repeated
coordinates.} 

\end{deluxetable}

The end result is a table (not shown) with 15 columns, one for each
catalog,\footnote{Although Table \ref{tab:master} lists 17 catalogs, two of
these refer to one of the other lists for object identification and are
therefore not counted in the number of cross-identified catalogs.} and 2887 
rows, each row representing the cross-identifications of a single object.
Table  \ref{tab:master} gives, for each catalog, the number of objects in the
HRC FOV (col. 4). Column 5 gives a measure of the relative
uncertainty in the catalog coordinates, $R_{95}$, representing the 95\% quantile of the object's
offsets from the stars of the reference catalog \citep{jon88}. Column 6 reports
the number of objects used in the latter calculation.  With typical values of
$R_{95} \sim 0.5^{\prime\prime}$, the number of uncertain or
ambiguous identifications is indeed small. The number of objects in each
catalog identified with an HRC X-ray source is given in column 7. In the next
section we discuss in detail the important issue of the reliability and
uniqueness of these HRC source identifications.

\subsection{X-ray source identification \label{sect:hrd_id}}

\begin{figure}
\centerline{\psfig{figure=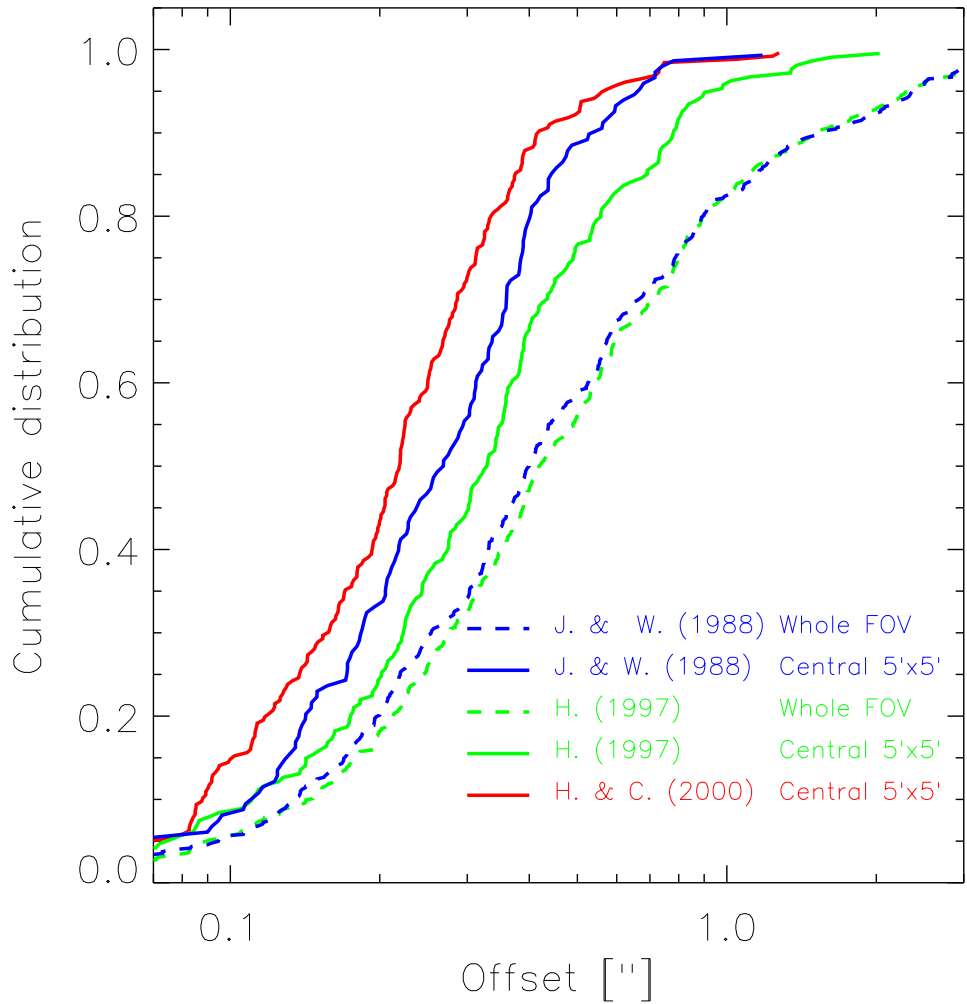,width=14cm}}
\caption{Cumulative distributions of offsets between X-ray and optical
positions, for three optical catalogues: \citet{jon88,hil97,hil00}.
For the first two, we show separate distributions for the inner
$5^{\prime} \times 5^{\prime}$ and for the entire FOV.  Since the
third catalog covers only the central region, a single
distribution is shown. \label{fig:offsets}}
\end{figure}

The outstanding sharpness of the {\em Chandra} PSF, especially in the field
center, makes it possible for the first time to identify detected X-ray sources
as easily as can be done in the optical and IR pass bands, even in an extremely
crowded field like the ONC. This fact largely eliminates the identification
uncertainties that were unavoidable with all previous X-ray observatories and
makes assignment of physical parameters for the X-ray emitters much easier than
before. Figure \ref{fig:offsets} shows the cumulative distribution of the
offsets of our HRC sources with respect to three of our component catalogs.
Offsets near the field center are particularly
small and depend systematically on the comparison catalog, the positions of
\citet{hil00} being most similar to {\em Chandra's}. It is therefore quite
likely that optical/IR uncertainties are an important (if not predominant)
contribution to the field-center offset distribution.
Although identification radii are small, the surface density of objects in our
catalog is extremely high, especially in the central $\sim 5^{\prime} \times
5^{\prime}$ Trapezium region where both the actual density of ONC members and
the sensitivity of the optical/IR catalogs are highest. Fortunately the field
center is also where the HRC PSF is sharpest.

In order to verify that the probability of spurious identification is small, we
calculated the expected number, $N_{ch.id}$, of chance correlations in a given
region of our image by estimating the object surface density at the positions
of our sources ($\rho_{obj}$), multiplying these densities by the coincidence
region areas\footnote{For this calculation, we took a quadrature sum of X-ray
position uncertainty and a generous optical uncertainty of
$0.75^{\prime\prime}$ as the identification radius.} ($A_{id}$) and then
summing over all sources, i.e.:

\begin{equation}
N_{ch.id}=\sum_{i}{A_{id}^i \cdot \rho_{obj}^i}
\end{equation}

\noindent
where the sum is extended over all sources in the region of interest. To
estimate the surface density $\rho_{obj}$, we characterised the FOV in two
different ways: as square tiles of $1^{\prime}$ on a side, and as a series of
concentric annuli centered on the X-ray field, with the radius of each annulus
$0.5^{\prime}$ greater than that of the previous one. We then counted the
number of cataloged sources in each of these regions and divided by their area.
The two methods gave equivalent results. The probability of chance
identification is highest in the field center, $\sim 5\%$ for the inner
$1^{\prime}$ radius; for the entire field, the chance probability is $\sim2\%$. If anything, these probabilities could even be smaller, since actual identification radii for most
optical catalogues were smaller than the conservative $0.75^{\prime\prime}$
used here.

We can compare these estimates with the number of spurious X-ray sources
expected on the basis of the SNR threshold chosen for detection (cf.  \S
\ref{sect:detections}). With 10 such spurious detections expected in the full
FOV, none (i.e., $\lesssim 0.2$) will be associated by chance with a known
object.  This means we needn't worry about spurious sources when studying the
properties of previously cataloged objects.  When exploring the nature of
unidentified sources (cf. Paper~II), however, we must remember that $\sim
10$ detections will be spurious.

We can also estimate the number of times an X-ray source identified with its
true counterpart will be identified by chance with a second object. Assuming
that the object positions are uncorrelated, we multiply the fraction of chance
identifications (2\%) by the total number of sources (742) to get an expected
$\lesssim 15$ double chance identifications in the entire field.\footnote{With
chance double identifications scaling as the square of the radius, and for
X-ray errors of $\gtrsim 2^{\prime\prime}$ typical of {\em XMM-Newton} and the
best prior studies \citep{gag95a}, the expected fraction of double ID's in the
central $1^{\prime}$ would be $\gtrsim 45\%$ (for {\em Chandra}-level
sensitivity and assuming uncorrelated object positions). Even neglecting source
confusion problems, no other past or present X-ray telescope is suitable for
ONC study.}  Since we actually find 50 multiple ID's, the source positions probably are correlated.

\subsection{Physical properties of cataloged stars \label{sect:phys_prop}}

For each object in our merged catalog we collected relevant data such as
X-ray count rates, optical and IR photometry, rotational periods,
etc. In cases of redundant information, only one of the available values was
chosen: the last column of Table \ref{tab:master} reports the information used from
each catalog, and in cases of redundant 
information, a superscript on a quantity indicates the precedence rank with which values were adopted for our
merged list (e.g., $H^5$ was used only if the other four catalogues with $H$ values had no entry for a given object).  Table \ref{tab:opt_data} gives the number of objects in our
catalog for which we retrieved several measured or estimated quantities, as
well as the same number referred to HRC sources identified with a single
optical/IR object.
The most complete data for objects are the near-IR $H$ and $K$ 
magnitudes, since the majority of the ONC population 
is embedded in the molecular cloud and is only detectable at IR wavelengths
where the extinction ($A_K \sim 0.1 A_V$) is greatly reduced from that in the optical.

\begin{figure}
\centerline{\psfig{figure=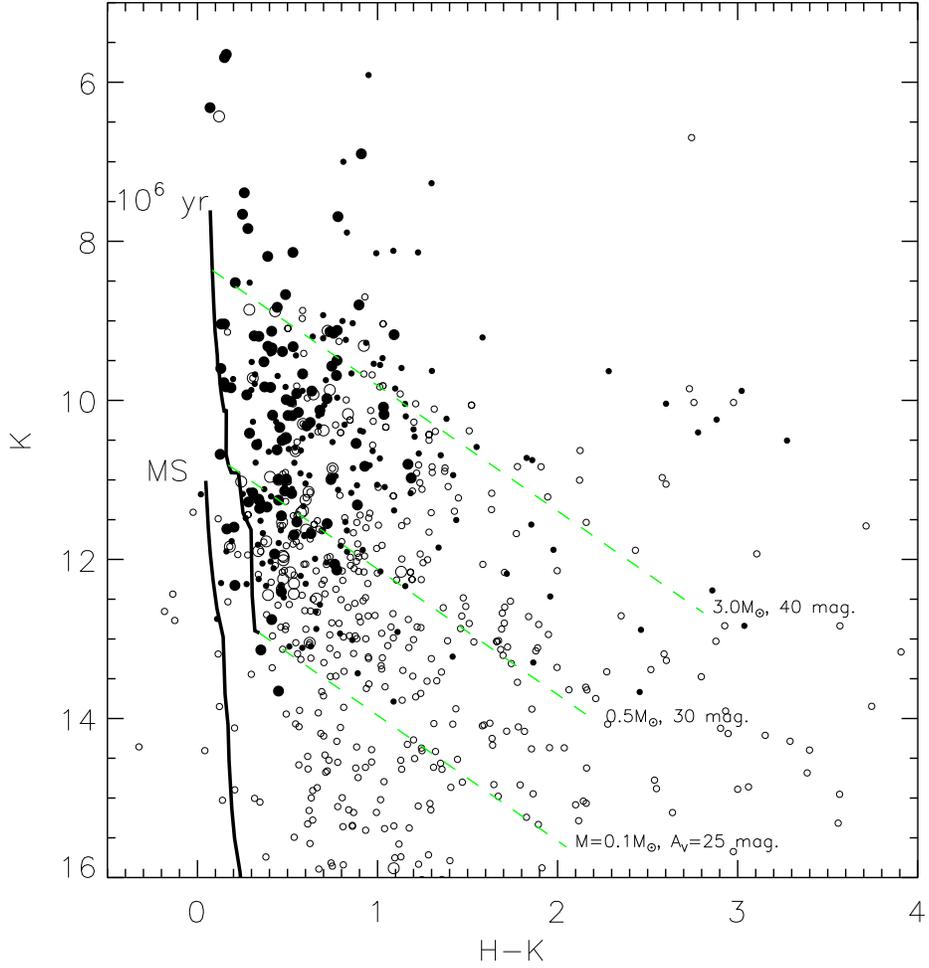,width=14cm}}
\caption{Near-IR color magnitude diagrams for the central $5^{\prime} \times 5^{\prime}$
of the HRC FOV. Large and small symbols indicate stars belonging and not belonging
to the {\em optical sample} (see \S \ref{sect:sample}). Filled and empty symbols indicate objects unambiguously
detected and not detected, respectively, in the HRC data. Also shown are the main sequence and the 1 Myr isochrone
according to SDF evolutionary tracks. \label{fig:khk}}
\end{figure}

A first look at our stars via
IR photometry is shown in Figure \ref{fig:khk} as a $K$ vs. $H-K$ diagram for the 763
stars in the inner $5^{\prime}\times5^{\prime}$ of the FOV.
The same diagram for the entire HRC FOV would be similar but, with 2431 stars, it would be overcrowded. X-ray sources are depicted with
filled symbols and optically well characterized stars
are shown with larger, open circles. Also shown are the
main sequence and a 1 Myr. isochrone derived by transforming the
evolutionary tracks of \citet[][ hereafter SDF]{sie00} from the theoretical ($L_{bol}$ vs. $T_{eff}$) to the
observational plane using the transformations of \citet{bes91} and \citet{hou00}.
Reddening vectors for 1 Myr stars of three different masses are indicated by
dashed lines. We cannot derive precise information on the nature of our stars
from this diagram alone: $H$ and $K$ band magnitudes are indeed expected to be
influenced  by the presence of disks around many of our stars. We expect that
stars with disk will be displaced with respect to those without disks toward
the upper right of Figure \ref{fig:khk}. For 813 stars in common with our sample, \citet{hil98a} derive a mean
\footnote{The corresponding median value is 0.26 mag.}
$I_c-K$ color excess of 0.36 magnitudes, with a standard deviation
of 0.72 mag. Given that disk
emission is expected to be minimum at $I$ band wavelengths, these values can be
taken as a rough measure of the average $K$ band excess and as an upper limit
to the $H-K$ color excess. We can conclude that the positions of stars in
Figure \ref{fig:khk}, although influenced by excesses, are approximately
representative of their photospheres. We then observe that our near IR
catalog contains stars down to very low masses ($M << 0.1M_\odot$) and  high
extinction ($A_V > 50$), and that stars are detected in our HRC data down to $M
\gtrsim 0.1 M_\odot$ and to quite high extinction ($A_V \sim 40$).

\begin{deluxetable}{lcccccccc}
\tabletypesize{\scriptsize}
\tablecaption{\small Numbers of objects with each optical/IR data item\label{tab:opt_data}}
\tablewidth{0pt}
\tablehead{
\colhead{ } & 
\colhead{$Spec.~~Type$} & 
\colhead{$A_V$} &
\colhead{$L_{bol}$} &
\colhead{$T_{eff}$} &
\colhead{$Mass$} &
\colhead{$Age$} &
\colhead{$P_{rot}$} &
\colhead{$Memb$\tablenotemark{\dag}} 
}
\startdata
Whole sample	& 1001 & 924 & 907 & 908 & 900 & 873 & 451 & 1028 \\
HRC detections  &  460 & 437 & 437 & 437 & 434 & 427 & 242 &  492 \\
\tableline\tableline
		 &	&     &     &	  &	&     &     &	   \\
		&$\Delta_{I-K}$&$EW_{CaII}$&$V$&$I$&$J$&$H$&$K$&$L$ \\
		 &	&     &     &	  &	&     &     &	   \\
\tableline
		 &	&     &     &	  &	&     &     &	   \\
Whole sample     &   785&  872& 1176& 1625& 2339& 2532& 2466&	132\\
HRC detections   &   389&  385&  501&  596&  604&  642&	641&	 49\\
\enddata
\tablenotetext{\dag}{Indicates stars with measured proper motion.}
\end{deluxetable}

One of our main goals, to be pursued further in Paper~II, is to study the activity of young 
stars as a function of their mass and evolutionary stage. The most relevant 
data in this context are those given in the papers by \citet{hil97} and 
\citet{hil98a}. The largest part of our spectral types, extinction estimates
($A_V$), $T_{eff}$ and $L_{bol}$ measurements are taken from \citet{hil97}, and
we refer the reader to that work  for details. The same author also derives
stellar masses and ages using the \citet{dan94} evolutionary tracks. Since new
improved model calculations have been published, we decided to estimate masses
and ages anew from the published temperatures and luminosities. We considered
two sets of calculations: those by \citet[][ hereafter DM97]{dan97}\footnote{The authors issued
an update in 1998 for $M < 0.9M_{\odot}$, available on-line at: http://www.mporzio.astro.it/$\sim$dantona/prems.html}  ($Z=0.02$, $Y=0.28$,  $X_d=2
\times 10^{-5}$) and
those of SDF ($Z = 0.02$, $Y = 0.277$, $X = 0.703$ and no
overshooting). Recent estimates of dynamical masses for seven
single stars and two binary systems by \citet{sim00} seem to favor the SDF 
over the DM97 tracks. We also observe that for ONC members the SDF tracks
produce a smaller age spread than do the DM97 calculations, possibly
better reflecting the real star formation history. Given uncertainties in
the models, it is premature to rely exclusively on a particular set of tracks. We
primarily have adopted the SDF tracks and in Paper~II will note whether and
how our results are influenced by this choice over that of the DM97
calculations.

Masses and visual absorptions for 19 presumed brown dwarfs ($0.008 \le
M/M_{\odot} \le 0.08$) and 3 low mass stars studied with IR spectroscopy by \citet{luc01}
were used when available. For four of these objects we could also
derive masses using the optical data of \citet{hil97} and the SDF tracks: 
three are brown dwarfs ($M_{IR}=$ 0.073, 0.9074 and 0.071  $M_\odot$) according to the IR
data and very low mass stars ($M_{Opt.}=$ 0.11, 0.11 and 0.19 $M_\odot$) according to the
optical estimate; the fourth is a very low mass star with $M_{IR}=0.3M_\odot$
and  $M_{Opt.}=0.21M_\odot$.



\subsection{Definition of stellar samples \label{sect:sample}}

About a third (907) of the cataloged objects can be placed on the HR diagram
and belong to the well characterized population studied by \citet{hil97} and
\citet{hil98a}. We were
able to estimate masses and ages for 877 of these stars that lie within the SDF tracks.
Masses for nine massive stars ($M/M_\odot
> 6.0$) were taken from \citet{hil97}. For four stars in the SDF tracks (see
\S \ref{sect:phys_prop})
and for 18 other very low mass stars and brown
dwarfs, we adopt values by \citet{luc01}. For the most part, the remaining
cataloged objects ($\sim 2000$) are heavily obscured stars detected only in the
near IR and for which we therefore have limited information. 

\citet{hil97} argues that her sample, which comprises most of our well
characterized sample, is representative of the whole ONC population and differs
only in that its location on the near side of the molecular cloud gives it
low optical extinction. This is important for our study of X-ray
activity (Paper~II) because it ensures that our results will not be strongly
biased.  As expected in a flux limited sample, however, the
brighter and more massive stars studied by \citet{hil97} are actually seen to
larger values of visual extinction than their lower mass
counterparts. In order to minimize selection effects for our studies,
we limited part of our analysis to stars with $A_V < 3.0$. 

For the following analysis we define two groups of stars, mainly on the
basis of the amount of available information: 

1) Our {\em optical sample} is comprised of stars in the HRC FOV for which we have a
mass estimate, whose values of $A_V$ are less than 3.0, and which are either confirmed
proper motion members or have unknown proper motion (cf. \S 
\ref{sect:cont_opt}). For the 696 stars of this {\em optical sample}, Table \ref{tab:opt_sample} lists:
sky position, mass, age,
rotational period, Ca II line equivalent width, HRC {\em basal} count rate (see \S \ref{sect:Xprop}),
X-ray luminosity and $L_X/L_{bol}$  (cf.
\S  \ref{sect:optIRprop} and \S \ref{sect:Xprop}).  The analysis
of X-ray activity that we present in Paper~II is based on these data.

\begin{deluxetable}{rcccccrrrr}
\tabletypesize{\tiny}
\tablewidth{0pt}
\tablecaption{Catalog of the ONC {\em optical sample} \label{tab:opt_sample}}
\tablehead{
\colhead{$N.$}&       \colhead{$Ra (2000)$}&   \colhead{$Dec (2000)$}&
\colhead{$Mass$}&     \colhead{$Log(Age)$} &   \colhead{$P_{rot}$}   &   
\colhead{$EW(CaII)$}& \colhead{$Ct. Rate$\tablenotemark{a}} &   \colhead{$Log(L_X)$}  &   
\colhead{$Log(L_X/L_{bol})$}   \\
\colhead{}&   \colhead{[$~h$ $~m$ $~s$]}&   \colhead{[$^\circ$ \ $^{\prime}$ \ $^{\prime\prime}$]}&
\colhead{$[M_\odot]$}&   \colhead{[yr.]}&   \colhead{[days]}&   
\colhead{}&  \colhead{[ks$^{-1}$]}&  \colhead{$[ergs\cdot s^{-1}$]}&   \colhead{}\\
\colhead{(1)}&
\colhead{(2)}&
\colhead{(3)}&
\colhead{(4)}&
\colhead{(5)}&
\colhead{(6)}&
\colhead{(7)}&
\colhead{(8)}&
\colhead{(9)}&
\colhead{(10)}
}
\startdata
  1&     5    34 12.81 &    -5    28 48.28 &      0.72&      6.46&    \ldots&      1.60&  $<$      4.02&  $<$  30.24&  $<$     -3.14\\
  2&     5    34 14.39 &    -5    28 16.77 &      1.74&      6.87&    \ldots&      2.00&          27.39&       31.21&          -3.12\\
  3&     5    34 17.15 &    -5    29 04.45 &      0.23&      4.71&    \ldots&      0.00&  $<$      3.28&  $<$  30.57&  $<$     -3.08\\
  4&     5    34 17.92 &    -5    33 33.45 &      0.43&      6.04&      3.19&      1.80&           4.59&       30.23&          -3.23\\
  5&     5    34 19.39 &    -5    27 12.04 &      1.86&      6.79&    \ldots&      4.00&  $<$      2.83&  $<$  30.04&  $<$     -4.35\\
  6&     5    34 19.47 &    -5    30 19.99 &      1.19&      6.20&    \ldots&     -2.50&  $<$      3.62&  $<$  30.48&  $<$     -3.42\\
  7&     5    34 20.64 &    -5    32 35.24 &      0.19&      6.29&    \ldots&      1.80&  $<$      3.75&  $<$  30.14&  $<$     -2.70\\
  8&     5    34 20.72 &    -5    23 29.18 &      0.28&      6.55&    \ldots&      0.00&           1.74&       29.81&          -2.96\\
  9&     5    34 20.92 &    -5    24 48.53 &      0.25&      6.52&    \ldots&      0.70&  $<$      2.56&  $<$  29.97&  $<$     -2.78\\
 10&     5    34 22.32 &    -5    22 27.01 &      0.12&      5.07&    \ldots&    \ldots&  $<$      2.40&  $<$  29.95&  $<$     -3.08\\
\enddata

\tablenotetext{a}{{\em Basal} HRC count rate (see text)}
\tablecomments{ The 696 entries of the full table are available on-line at:
http://www.astropa.unipa.it/$\sim$ettoref/PaperI\_tables.html}
\end{deluxetable}

2) The {\em IR sample} is comprised of 2476 stars with measured $K$ band magnitudes.
The {\em IR sample} includes most of the {\em optical sample}, with 680 of 696
stars in common.  This sample very likely contains a large fraction of all the
ONC members of stellar and brown dwarf mass, with the possible exception of a
few deeply reddened stars located in small regions where the cloud absorbing
column is greatest (see Paper~II).  The central region, which also coincides
with a thicker part of the molecular cloud, has been surveyed in depth by 
\citet{hil00} and \citet{luc00}. The survey of \citet{hil00} is believed 
better than 90\% complete to $K \sim 17.5$,
corresponding to a mass of $\sim 0.02
M_{\odot}$ (for an age of 1 Myr).

\begin{figure}
\centerline{\psfig{figure=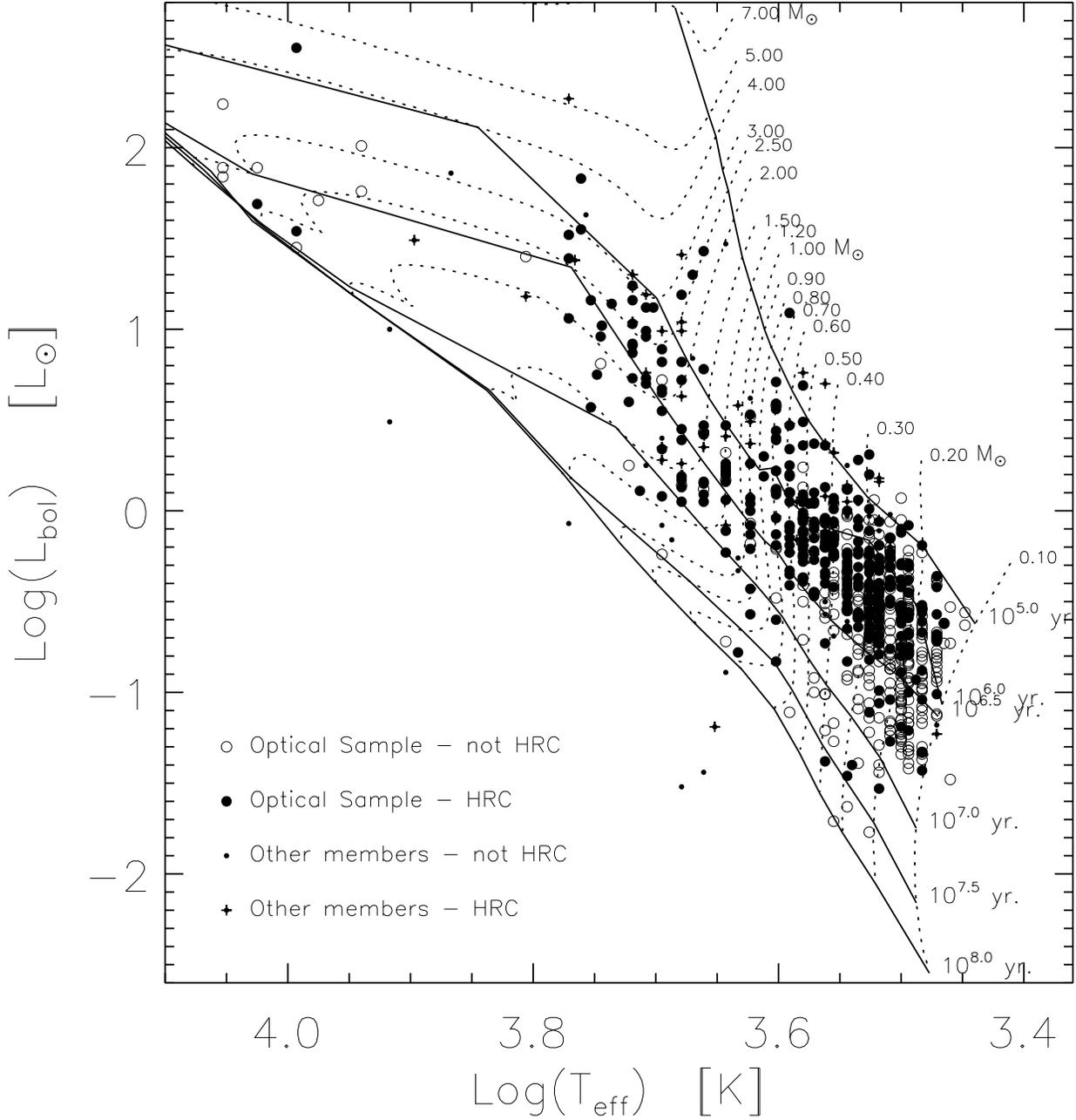,width=18cm}}
\caption{HR diagrams for stars belonging to the {\em optical sample} and for other ONC members which we could place in the diagram; see legend for symbol meanings.
We indicate with filled circles and small crosses the stars in these two groups that were detected in the HRC image. Dashed and solid lines, respectively, represent SDF evolutionary tracks for masses from 0.1 to $7.0M_\odot$ and isochrones for ages between $10^5$ and $10^8$ years (note labels besides curves). \label{fig:hr_sdf}}
\end{figure}

Figure \ref{fig:hr_sdf} shows an HR diagram for the stars in our {\em optical sample}.
We also show the positions of other
confirmed proper motion members ($P_{memb}> 50\%$) which are not part of the
{\em optical sample}. For both groups we indicate stars (cf. figure legend)
that are unambiguously identified with our X-ray sources. SDF tracks and
isochrones used to estimate stellar masses and ages are also shown for
reference.

In the following two sections we estimate the
degree to which our two main stellar samples are contaminated by field objects not
physically or spatially related to the ONC. 

\subsubsection{Contamination of the optical sample \label{sect:cont_opt}}

In discussing the {\em optical sample} and its HR diagram we will make use of
the membership information provided by \citet{jon88}. We consider
three stellar subsamples that differ in confidence of
their physical association with the ONC. The first, 
``{\em group 0},'' comprises all stars studied
astrometrically by \citet{jon88}, regardless of membership probability
as derived in that study.  The next sample, ``{\em group 1},'' comprises all stars
with proper-motion membership probabilities greater than 50\%.  Thirdly, we
consider the {\em optical sample} defined in the previous section, which
excludes only those stars explicitly
suspected of being non-members ($P_{memb} < 50\%$).

We can estimate the number of field objects, $N_{field}$, that contaminates a
(sub)sample of stars studied astrometrically \citep{jon88} by
interpreting proper motion probabilities literally: $N_{field}=\sum (1.0-P_i)$, where
$P_i$ is the probability that the $i^{th}$ object belongs to the ONC and the
sum is extended over the whole (sub)sample. Table \ref{tab:field} displays our results, with
the first two rows referring to stars in the
entire HRC FOV and the inner $5^{\prime} \times 5^{\prime}$ region. The
remainder of the table reports the full-field analysis
as a function of stellar mass and age.
The table gives total numbers of objects and field contamination fractions estimated
for the stars in groups 0 ($P_{memb}$ known; columns 2 and 3) and group 1 ($P_{memb} \geq 50
\%$; columns 4 and 5).  The last two columns refer to the {\em optical sample}, which
contains both stars with $P_{memb} \geq 50 \%$ and stars with no proper
motion information. For the former stars we estimate the fraction of field objects
using the same procedure as for groups 0 and 1; for the latter, we assume that
this fraction is the same as for group 0 (i.e., we use the values from column 3). 

Regarding sample contaminations, we therefore draw the following conclusions: a) Field contamination for the
whole sample is a factor of $\sim 2$ lower ($\sim 5\%$) in the central region
with respect to the full FOV ($\sim 10\%$), most likely due to the high member
surface density and to the optically opaque molecular cloud that effectively
obscures background stars. b) Selection of stars with $P_{memb} \geq 50 \%$
(i.e., group 1) reduces the contamination to $2-2.5\%$ regardless of the area
considered. c) Contamination depends on position in the HR diagram, being
highest for masses between 1 and 10$M_{\odot}$ and for the oldest stars
($Log(age) > 6.5$) in group 0; 
this is likely the result of both the distribution of field stars in the observational HR diagram and of the dependence on mass and age of the member sky density (the $3-10M_\odot$ stars, for example, are the most uniformly distributed among the mass ranges considered and, given the centrally peaked member distribution, are found in regions of higher field star relative density).
d) Contamination in the {\em optical
sample} is probably not much larger than in group 1, our estimate being $3.5\%$
for the full FOV and $2.4\%$ for the central region. We therefore take
the {\em optical sample} as our primary basis for the analysis of Paper~II.

\begin{deluxetable}{lcccccc}
\tabletypesize{\scriptsize}
\tablecaption{Field star contamination for various samples\label{tab:field}}
\tablewidth{0pt}
\tablehead{
\colhead{Sample} & 
\colhead{N(0)\tablenotemark{a}}& 
\colhead{Field \% (0)}& 
\colhead{N(1)\tablenotemark{b}}& 
\colhead{Field \% (1)}& 
\colhead{N(Opt)\tablenotemark{c}} &    
\colhead{Field \% (Opt)\tablenotemark{d}} \\
\colhead{(1)}&{(2)}&{(3)}&{(4)}&{(5)}&{(6)}&{(7)}}
\startdata
Whole FOV                         & 1034 & 10.9 & 938 & 2.4 & 696 & 3.5 \\
Inner $5^{\prime}\times5^{\prime}$&  241 &  4.6 & 235 & 2.2 & 138 & 2.4 \\
                                  &      &      &     &     &     &     \\
\tableline
               & \multicolumn{4}{c}{Mass and age breakdown} &     &     \\
\tableline
                                  &      &      &     &     &     &     \\
$M/M_{\odot}=  0.10-  0.16$ &    43&  4.5&    42&  2.2&       61&  2.9\\
$M/M_{\odot}=  0.16-  0.25$ &   135&  5.6&   129&  1.6&      160&  2.4\\
$M/M_{\odot}=  0.25-  0.50$ &   302&  5.1&   291&  1.8&      277&  2.2\\
$M/M_{\odot}=  0.50-  1.00$ &   115&  9.8&   106&  2.5&       89&  2.8\\
$M/M_{\odot}=  1.00-  2.00$ &    76& 26.5&    57&  3.5&       40&  5.1\\
$M/M_{\odot}=  2.00-  3.00$ &    53& 14.9&    47&  4.3&       37&  4.7\\
$M/M_{\odot}=  3.00- 10.00$ &    23& 24.1&    18&  5.4&       13&  6.5\\
$M/M_{\odot}= 10.00- 50.00$ &     6&  5.3&     6&  5.3&        6&  5.3\\
$Log(age)=  4.50-  5.50$    &    61&  5.3&    59&  2.1&       56&  2.6\\
$Log(age)=  5.50-  6.00$    &   136&  8.4&   126&  1.9&      117&  2.6\\
$Log(age)=  6.00-  6.50$    &   366&  5.6&   352&  2.3&      330&  2.5\\
$Log(age)=  6.50-  7.50$    &   167& 18.7&   139&  2.7&      162&  6.4\\  
\enddata
\tablenotetext{a}{Number of stars with known proper motion \citep{jon88}.}
\tablenotetext{b}{Number of stars with $P_{memb} \geq 50\%$} 
\tablenotetext{c}{Number of stars in the {\em optical sample}}
\tablenotetext{d}{Extrapolated fraction of field contamination (see text).}
\end{deluxetable}

\subsubsection{Contamination of the IR sample}

For stars detected only in the $K$ band, we cannot perform the above estimates
because of lack of proper motions or any other individual membership data.
Nonetheless it is particularly important to have at least a statistical understanding of the
contamination, since we expect deep $K$ band observations to
begin to see through the dense molecular cloud.

We estimate the expected contamination by field objects (galactic and
extra-galactic) with the following procedure. Briefly, we first determine a reasonable
approximation to field objects surface density as a function of unabsorbed $K$
magnitude. Since the majority of field objects will be in the background 
with respect to the ONC, their
density at a given magnitude will be lowered by absorption in the molecular cloud. A smaller, unabsorbed fraction will be in
the foreground. We will assume that the shape of the
$K$ distribution for the two populations is the same as that of the total field
population. Second, in order to know the total number of contaminating non-members,
we must estimate the ratio of the number of foreground to background stars.
Thirdly, we use the  $^{13}CO$ measurements published by
\citet{gol97} to estimate the total molecular cloud absorption in $K$ ($A_K$)
as a function of sky position. Note that because these data are only available in the field center the contamination can only be estimated in the central region. Fourth, we convolve the background star $K$
distribution with the $A_K$ distribution over the area of interest,
and fifth, we add the unmodified foreground star distribution to
obtain the expected density of the field stars as a function of $K$.

In the first step of the calculation outlined above, the field star density as a function of $K$ was derived from more than 13600
stars in the 2MASS survey\footnote{see http://www.ipac.caltech.edu/2mass/} 
within $1^\circ$ of the Trapezium: although this distribution is affected by the
presence of the cluster and of the absorbing cloud, the vast majority of stars
from which it is derived come from the field and are located off the cloud so
that the final result will not be significantly affected. It is important to
realize what the limiting magnitude of our $K$ band luminosity function is. 
Although the nominal 2MASS survey completeness limit in $K$ is 14.3, in many
cases it goes fainter than this, and for the $1^\circ$-radius region considered
here, the $K$ band luminosity function rises to $K \sim 15$ and turns over for
fainter $K$ magnitudes. We will therefore take $K=15$ as our limit to the
knowledge of the $K$ distribution. Considering that more than 90\% of the area
of interest has a  $^{13}CO$ column density corresponding to $A_K > 1.0$, we
can confidently estimate background contamination down to $K \sim 16$.

In the second step above, where we estimate the fraction of foreground stars
in the direction of the ONC and in a
given solid angle, $f_{ONC}=N_{ONC}(d<470pc)/N_{ONC}(tot)$, we use the results
obtained by \citet{car00} through the \citet{wai92} galactic model. Although
the author does not report a result for the ONC, he quotes a ratio of
foreground to total field stars for Perseus, $f_{Per}=N_{Per}(d<320 {\rm
pc})/N_{Per}(tot)$, which has a similar galactic latitude ($b\sim -20^\circ$) to
that of the ONC and a distance of 320 pc (vs. 470 pc). We assume that the
object space density as a function of distance, $\rho(d)$, is the same in the
two directions: with this assumption the fraction of foreground stars in the
ONC can be expressed as:

\begin{equation}
\label{eq:field_fract}
f_{ONC}=f_{Per}\cdot \frac{\langle{\rho(470)}\rangle}{\langle{\rho(320)}\rangle} \cdot \left(\frac{470}{320}\right)^3 \lesssim f_{Per}\cdot \left(\frac{470}{320}\right)^3
\end{equation}

\noindent
where $\langle\rho(d)\rangle$ is the mean stellar density within a distance $d$
and the last inequality is justified by the larger height of the ONC above the
galactic plane. According to \citet{car00} $f_{Per}$ ranges from 0.05 to 0.08
in the $J$, $H$ and $K$ bands. We have (conservatively) taken the upper end of
this range and scaled it by cube of the distances (eq. \ref{eq:field_fract}),
therefore deriving $f_{ONC} \lesssim 0.25$. We will conservatively assume in
the following that $f_{ONC}=0.25$; this could result in an overestimation of field
contamination.

The fifth step of our calculation gives us the expected surface density of field
stars in the central $5^{\prime}\times 5^{\prime}$ region as a function of
magnitude. For example, we find a value of 0.16 stars ${\rm arcmin}^{-2}$ for
$K<13$ and 0.44 stars ${\rm arcmin}^{-2}$ for $K<15$. These numbers can be
compared with the observed mean densities in the same region at the same two
limiting magnitudes: $\sim 21$ and $\sim 28$ stars ${\rm arcmin}^{-2}$  for $K
< 13$ and $K < 15$ respectively. Even for $K < 16$ the expected contamination
from field stars is about 2\%.

Having estimated the field star surface densities, we now compare our
determinations with others, performed with similar data and in a similar fashion.
Consider the densities of unabsorbed field
stars. \citet{car01}, in their work on IR variability, derive an {\em off-cloud}
density of 0.66 stars per square arcmin for $K < 14.8$.
\citet{car00} performs a similar calculation of the unabsorbed field star
densities for the ONC and other clouds. From the author's
Figure 3, we can see that the mean field star
density  for $K < 14.3$ at the galactic latitude of the ONC (i.e., -19 degrees)
is about 0.5. The densities we derive are: 0.81 for $K < 14.8$ and 0.64 for $K
< 14.3$, both within  20-30\% of the previous derivations. Because our goal is
to show that contamination is not significant, a more conservative approach is to adopt our slightly larger
derivation for the stellar surface density. 
Comparing our expectation for the density of observed field stars with the
results of \citet{hil00}, we find that their densities are about 50-60\% lower
than ours, in keeping with our conservative approach.

We conclude, therefore, that contamination in our IR sample is negligible in the
center of the FOV down to $K \sim 16$. Note that this conclusion might not hold
for the whole area under study, because of the lower average member densities
and thinner background molecular cloud. The IR photometry in the outer regions
is not as deep as in the center, however, so that even there, contamination is
possibly not significant.

\subsection{The BN/KL region \label{sect:BNKL}}

The BN/KL region, about $\sim 1^\prime$ northwest of the Trapezium, in the
densest part of Orion Nebula Cloud, contains a cluster of massive embedded
stars mostly observed in the IR. The X-ray emission of the IR sources in the
region has been already investigated by \citet{gar00} using {\em Chandra}
ACIS-I data. We estimate, using PIMMS,\footnote{Portable, Interactive,
Multi-Mission Simulator, version 3.0, available on-line at:
http://asc.harvard.edu/toolkit/pimms.jsp} that our HRC-I data is 1.5 to 4.0
times less sensitive, depending on the source $kT$ and $N_H$ (see \S
\ref{sect:conv_fact}), with the worst case corresponding to high absorptions
($N_H \sim 10^{23}$). With reference to the sky region depicted in Figure 6 of
\citet{gar00}, our master catalog contains 61 distinct objects in the $\sim
0.6$ arcmin$^2$ area, 22 of which are detected by the HRC vs. 27 by ACIS (Table
2 in \citealt{gar00}). Out of the five ACIS sources missed by the HRC three are
classified as IR sources and two as X-ray only sources by \citet{gar00}. We do
not detect the BN object itself, but our upper limit ($< 12$ photons in 63
HRC-I ksec) is consistent with the low measured ACIS-I count-rate (18 photons
in 48 ACIS-I ksec). The IR source "n" (our source 272) likely varied between
the time of the two observations: the HRC detects 49 photons while ACIS detects
61, which, assuming $N_H=10^{23}$ \citep{gar00} and $kT > 1$ keV, indicates
that the HRC luminosity is $\gtrsim 3$ time greater than that measured by ACIS.
The IR source \#17 in Table 2 of \citet{gar00}, our source 281 in table
\ref{tab:detections} was, with  similar assumptions (but $N_H=10^{22.5}$), 6
times brighter in the HRC data. Two objects detected exclusively in X-rays,
also appear to have varied:  \#16 (HRC source 284) being $\sim  3$ times
brighter in the HRC;  and \# 23, undetected by the HRC and whose upper limit
indicates a variation of at least a factor of 3.

\section{Optical/IR properties of the HRC sources \label{sect:optIRprop}}

In this section we discuss the K magnitude and estimated mass composition of the counterparts of our X-ray source sample.
In doing so we often distinguish
the two samples of optically characterized members and near-IR objects.
In Paper~II, following our study of
correlations of X-ray activity with stellar parameters, we speculate on the
possible nature of X-ray sources not identified with optical or IR counterparts.

\begin{figure}[!t!]
\centerline{\psfig{figure=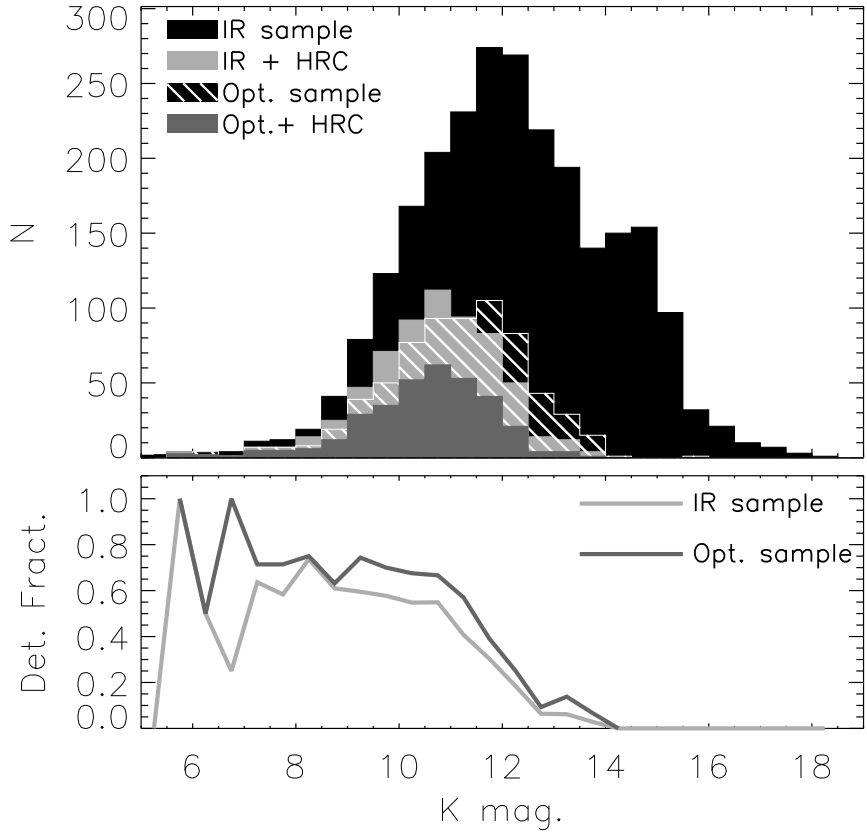,width=12cm}}
\caption{Upper panel shows K band distributions for: the {\em IR sample} (black), i.e., all objects within 
the HRC FOV with measured $K$ magnitudes; X-ray detected and unambiguously identified objects (light gray),
the well characterized {\em optical sample} (cross-hatched, see text); and {\em optical sample} objects detected in X rays (dark gray).
Lower panel displays HRC detection fractions for the {\em IR}
and {\em optical} samples (light and dark gray, respectively) as a function of $K$ mag.  \label{fig:k_dist}}
\end{figure}

The quantity most widely available for our cataloged objects is K band magnitude.
In the upper panel of Figure
\ref{fig:k_dist} we show the K band magnitude distribution for all our cataloged
objects together with that of the subsample of stars comprising our {\em optical sample}.
Also shown are distributions referring to objects in these two samples
that are detected in our HRC data and unambiguously identified with an IR or
optical counterpart. The lower panel of the figure shows detection fractions of the two
reference samples as a function of $K$. It can be seen that the fraction of
objects detected in X rays is greater than 50\% down to $K \sim 11$ but then
decreases for fainter $K$ magnitudes, going to zero at $K \sim 14$ which
roughly corresponds to the brown dwarf limit (cf. Figure \ref{fig:khk}) for
1~Myr-old unabsorbed stars. 
Similar results, although with slightly larger detection fractions, were obtained by \citet{gar00} with their ACIS-I data.
{\em Optical sample} stars follow a similar trend,
although with detection fractions somewhat larger.
A quantitative breakdown of detection fractions for several stellar samples and two limiting $K$ magnitudes is given in Table \ref{tab:det_break}. Note the increased detection fractions in the center of the FOV.

\begin{deluxetable}{llcc}
\tabletypesize{\small}
\tablecaption{Detection fractions vs. K magnitude\label{tab:det_break}}
\tablewidth{0pt}
\tablehead{
\colhead{}& Sky region &  $K \le 11$ & $K \le 14$
}
\startdata
{\em IR sample} 	& Full FOV    	& 0.57 & 0.32  \\
\hspace*{0.7cm} " 	& Central $5^\prime\times5^\prime$    	& 0.63 & 0.41  \\
{\em Optical sample} 	& Full FOV & 0.69 & 0.50  \\
\hspace*{0.7cm} "  	& Central $5^\prime\times5^\prime$ 	& 0.81 & 0.71  \\
{\em Optical sample} (max\tablenotemark{a} )& Central  $5^\prime\times5^\prime$ & 0.90 & 0.77  \\
\enddata
\tablenotetext{a}{Including ambiguous identifications}
\end{deluxetable}


\begin{figure}[!t!]
\centerline{\psfig{figure=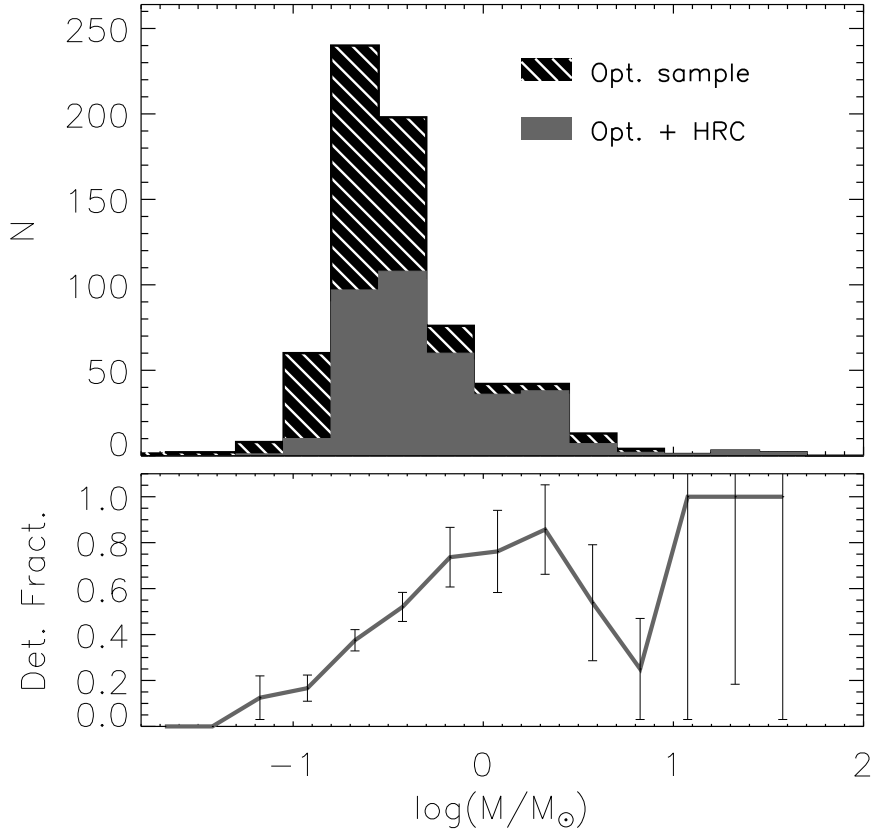,width=12cm}}
\caption{Upper panel: Mass distributions of objects in the HRC FOV and belonging to our {\em optical sample}.
Shadings same as in Figure \ref{fig:k_dist}. Lower panel: Detection fraction vs. mass; error bars
computed from Poisson statistics. \label{fig:M_dist}}
\end{figure}

Figure \ref{fig:M_dist} (upper panel) shows the distribution of stellar masses
for our {\em optical sample} and the subset of X-ray detected stars. As in the previous figure,
the lower panel shows the detection fraction: we see that the detection
fraction is 100\% for the most massive stars, drops abruptly for 
$Log(M/M_{\odot})=0.3-1.0$ ($M/M_{\odot} \sim 2.0-10.0$), rises again to about
90\% and then decreases fairly smoothly to zero in the brown dwarf regime.
A similar trend (not shown) is observed in the field center, although the detection
fractions are 10-20\% higher between $\sim 0.1$ and  $\sim 2.0M_{\odot}$.

%



\section{Determination of $L_X$ and $L_X/L_{bol}$ for HRC sources \label{sect:Xprop}}

Here we estimate two activity indicators, X-ray luminosity ($L_X$) and its
ratio to bolometric luminosity ($L_X/L_{bol}$), and estimate their associated
uncertainties. $L_X$ represents total coronal power output, while $L_X/L_{bol}$
is related to the fraction of total stellar energy that goes into heating the
corona. In our search for physical mechanisms that may drive activity in young
PMS stars,  Paper~II studies the relationship of these indicators with various
stellar parameters.

\begin{figure}
\centerline{\psfig{figure=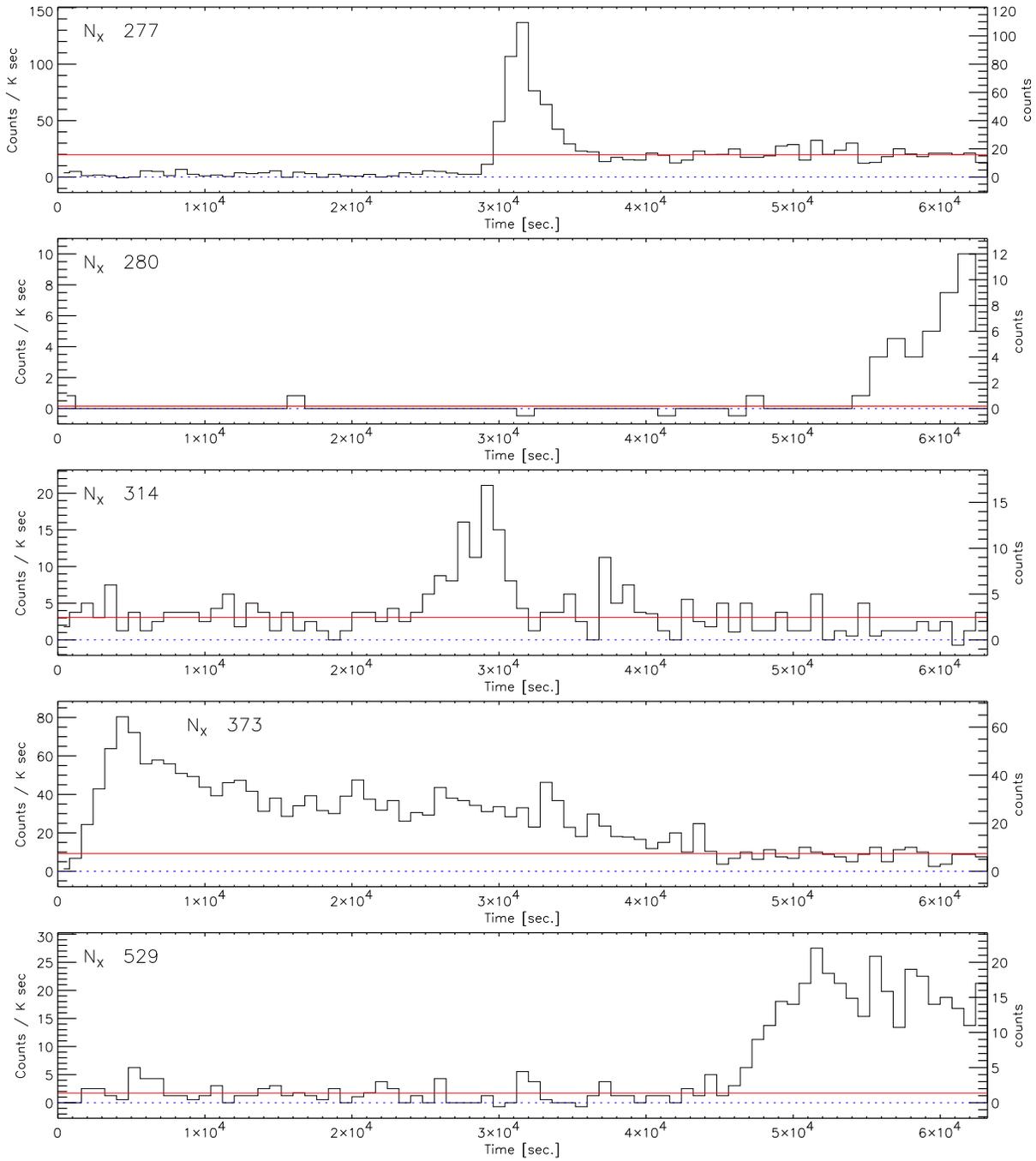,width=16cm}}
\caption{Light curves for five strongly variable HRC sources (number indicated at upper left 
of each panel) vs. time elapsed since observation start. Bin size varies from source to source,
and left and right ordinate scales show count rates and counts per bin, respectively.
Solid horizontal lines indicate derived {\em basal} rates (see text); dashed lines show zero 
levels. Source data were extracted from circular regions around each position, with subtracted background
estimated from larger concentric annuli. \label{fig:base_rate}}
\end{figure}

Because the intensity of many of the HRC sources varied during our observation,
we did not use mean count rates to estimate luminosities. Instead we have
defined a {\em basal} count rate to remove the effects of short term ($\lesssim
63$ ksec) variability. Our {\em basal} rate algorithm searches for that count
rate compatible with the largest possible portion of the light curve. Figure
\ref{fig:base_rate} shows five light curves of highly variable sources and
their {\em basal} rates, and Table \ref{tab:opt_sample} lists our estimates for
the basal count rates, $L_X$ and the ratio between $L_X$ and $L_{bol}$ for the
stars in our {\em optical sample}.

\subsection{{\em Chandra} HRC count rate to flux conversion factors (CFs) \label{sect:conv_fact}}

In order to convert {\em basal} count rates to intrinsic X-ray luminosity
(assuming a distance of 470 pc), we have derived conversion factors (CFs)
that yield unabsorbed source flux from observed count rate. 
We have assumed both that emitted radiation can be described as thermal,
optically thin, emission from an isothermal plasma at temperature
$kT$, and that the absorbing interstellar material
along the line of sight can be characterized by a hydrogen column density
$N_H$. We find that the X-ray flux of stars in our {\em optical sample} can be
adequately described by an isothermal ($kT = 2.16$) optically thin plasma,
absorbed by a gas column density proportional to optical extinction: $N_H= 2
\cdot 10^{21} A_V$.  This conclusion is based upon a spectral analysis of archival ONC X-ray data from
the Advanced CCD Imaging Spectrometer (ACIS, \citealt{tow00}), as discussed in
Appendix \ref{sect:ACIS_data}.
Because the {\em Chandra} HRC lacks spectral resolution, $N_H$ and $kT$ values cannot be deduced from our main dataset alone.

\begin{figure}[!t!]
\centerline{\psfig{file=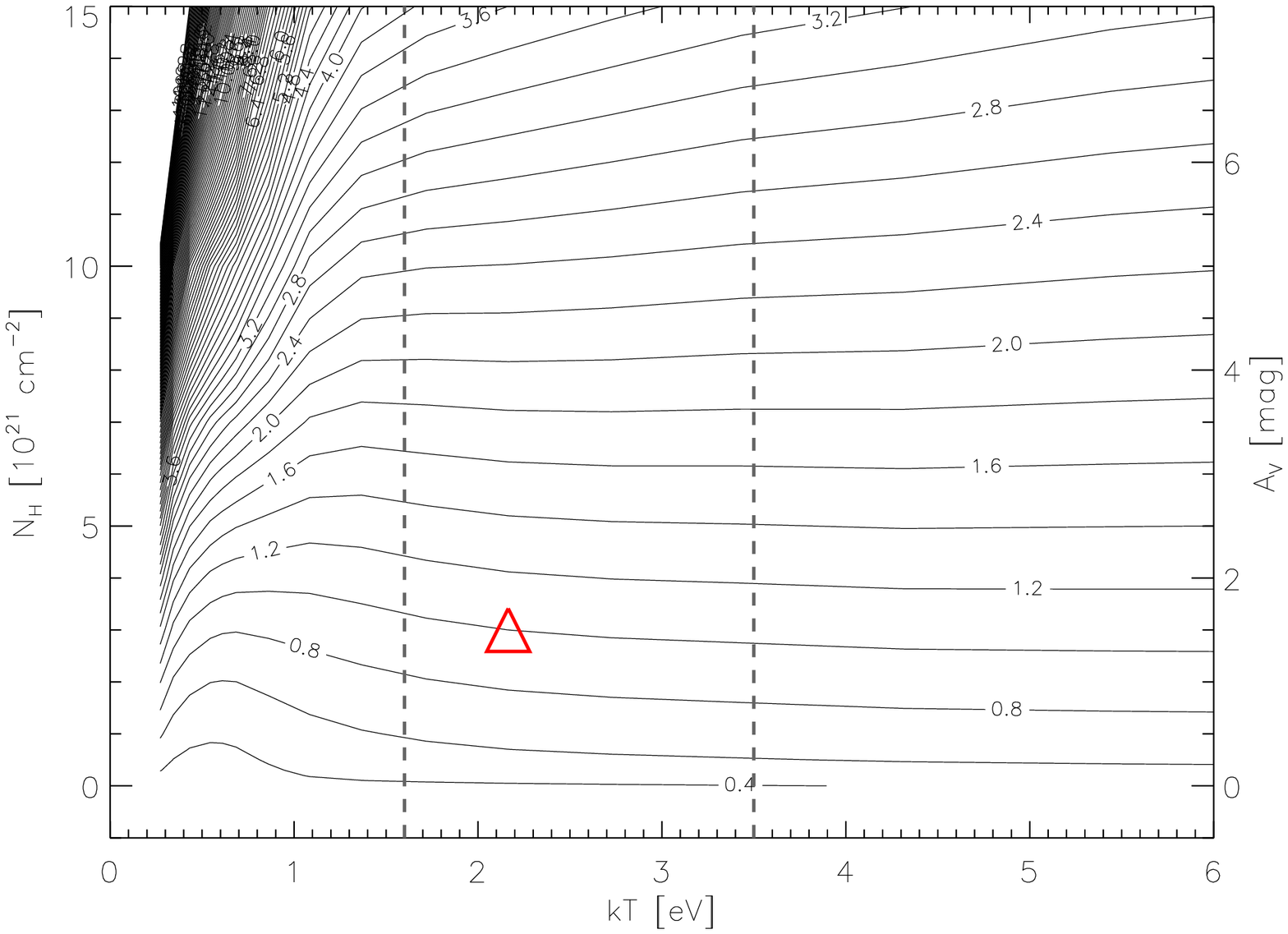,width=15cm}}
\caption{Contour plot of the ratio between the HRC conversion factor computed 
for $kT=2.16$ and $N_H=3\cdot 10^{21}$ ($2.9\cdot10^{-11} ergs\ cm^{-2}$ per photon; denoted by large triangle) and the same conversion factor at 
different values of $kT$ and $N_H$. Right vertical axis gives the {\em
standard} conversion between $N_H$ and $A_V$. Region between the two dashed
vertical lines depicts likely range of ONC source temperatures.
\label{fig:HRCcf}}
\end{figure}

In order to study the sensitivity of
the conversion factor to the assumed plasma temperature and hydrogen column
density we have explored the two dimensional ($kT$, $N_H$) parameter space in
the range $kT < 8.0$ and $20 < Log(N_H) < 23$ which most likely include the
conditions of the vast majority of our sources. All our HRC CFs are computed
using PIMMS,
a Raymond-Smith spectral model and a spectral band between 0.1 and 4.0
keV.\footnote{This band was chosen to ease comparisons with previous {\em
Einstein} and ROSAT results. A different choice would lead to different
conversion factors depending on source temperature: for the full {\em Chandra}
band (0.2-8.0) the fractional difference in CFs compared to our choice would
be between +7.5\% ($kT=1.3$) and -20\% ($kT=3.5$), with the most likely value
$\sim$ -4.5\% ($kT=2.16$).} Figure \ref{fig:HRCcf} shows a contour plot of the
ratio of the conversion factor at a given $kT$ and $N_H$ within the specified
ranges, to the conversion factor for typical ONC values of $kT=2.16$ and $N_H=
3\cdot 10^{21}$. The plot shows that the dependence of the conversion factor on
temperature and $N_H$ is quite weak for $kT > 1.0$ and values of $N_H$ around
the assumed value.  The weak dependence of the CF on temperature justifies (at
least for purposes of computing the CFs) the assumption of a single temperature
instead of the full EM distribution. Provided the majority of emitting
plasma is at temperatures larger than $\sim 1.0$ keV (likely for ONC PMS stars)
and the estimates of $N_H$ are reasonable, our approach will give useful results.

The next step, for each of our sources, is to estimate a temperature 
representative of the emitting plasma and a value of $N_H$ quantifying the
amount of interstellar gas between the source and the observer. If the range of
such representative temperatures is such that $kT > 1.0$ keV (cf. \S
\ref{sect:app_int}), we may safely use a single temperature ($\sim 2$ keV) to
compute the CFs for all sources. 
Although this $kT$ is somewhat higher than 1 keV, a value often used in past X-ray studies of PMS stars, the HRC conversion factors we thus derive differ, for typical $N_H$, by at most $0.1$ dex.
$N_H$, on the other hand, must be evaluated
individually since it certainly spans a large range of values.
We can estimate $N_H$ independently from
X-ray information by using the {\em standard}  linear correlation with the
optical absorption: $ N_H = 2.0 \cdot 10^{21} A_V$ 
\citep[cf.][]{ryt96}. Adopting such a linear relation between optical
absorption (due to dust grains) and gas column density
(responsible for X-ray absorption) assumes that the gas to
dust ratio is the same for all of our sources and equal to the value inferred
for field stars. This assumption might not hold for the ONC members for
two reasons: 1) most of the absorption is due to the molecular cloud in which
the cluster is embedded rather than to interstellar matter between us and the
cloud, and these two mediums might have different gas to dust ratios; and 2) the
circumstellar environment of these PMS stars, many of which are actively
accreting material from disks, is complex and might in principle contribute
to an anomalous extinction law. An independent check on the validity and limits
of the above assumption (see Appendix  \ref{sect:ACIS_data}) allows us to
proceed in using this $N_H$-$A_V$ relationship to derive $N_H$ from the $A_V$
values of \citet{hil97} for many of our ONC members. This is particularly
applicable for stars in our {\em optical sample} that enter into the
analysis of correlations between activity and stellar parameters (Paper~II).

To summarize this discussion, as well as that of Appendix \ref{sect:ACIS_data}, the
conversion between count rates and energy fluxes or luminosities is a crucial
step, usually requiring some spectral information.
We have satisfied this need with archival ACIS data and
conclude that our conversion factors are good to better than $\sim 30\%$ (0.1 dex), albeit
with a 1$\sigma$ scatter of $\sim 0.2$ dex. This conclusion does not appear to
depend on stellar mass or accretion rate.

\subsection{Uncertainties on $L_X$ and $L_X/L_{bol}$ \label{sect:act_unc}}

Uncertainties in our inferred X-ray luminosities are predominately due to the
conversion from counts to flux (except for the weakest of our sources, where
Poisson statistics are also important). In \S \ref{sect:conv_fact} we found
that the  statistical uncertainties on the CFs will be of the order of 0.2 dex
(1 $\sigma$). Count rate measurements have a median fractional Poisson
uncertainty of $\sim 17$\% (+0.07, -0.08 dex), smaller compared to the total
$L_X$ error, but for 5\% (or 35\%) of sources, this component is $\gtrsim 50$\%
(or 30\%), comparable to the conversion factor uncertainty. 

Distance uncertainties will also influence luminosity values.  At 470 pc, the
ONC's angular extent ($\sim 30^{\prime}$) corresponds to $\sim 4$ pc, and the
range of modern distance determinations introduces an additional systematic
uncertainty of $\sim 30$ pc, together yielding $\Delta L_X \approx 13\%$. Since
such largely systematic uncertainties won't qualitatively influence our results
for the dependence of coronal activity on stellar parameters, we neglect these.

The $L_X/L_{bol}$ ratio will be affected by uncertainties in both $L_X$ and
$L_{bol}$. For stars in our {\em optical sample} we have adopted the $L_{bol}$
values of \citet{hil97} with typical errors of $\lesssim 0.2$ dex (stars with
strong circumstellar accretion could have slightly larger, underestimation
errors). A quadrature sum of the uncertainties of $\sim 0.2$ dex for $Log(L_X)$
and 0.2 dex for $Log(L_{bol})$ yields a total $Log(L_X/L_{bol})$ uncertainty of
$\sim 0.3$ dex.

\section{Summary \label{sect:disc}}

Our {\sc pwdetect} wavelet algorithm finds 742 X-ray sources in the
$30^{\prime}\times 30^{\prime}$ FOV of our 63 ksec {\em Chandra} HRC-I
observation of the ONC.  As a basis for our study of ONC X-ray activity (see
Paper~II), we have combined 17 different catalogs from the recent literature to
assemble a catalog of all available information for nearly 2900 objects, the
large majority of which are ONC members.  We then calculated mass and age for a
significant subset using the evolutionary tracks of \citet{sie00} and defined
two reference stellar samples: the {\em optical sample}, comprising $\sim 700$
well characterized members with low extinction ($A_V \le 3.0$), and the
{\em IR sample}, largely including the {\em optical sample} and comprising
$\sim 2500$ stars observed in the $K$ band.  We have shown that field object
contamination is quite limited in both of these samples.

In order to characterize the population of X-ray detected objects, we have
presented distributions of $K$ magnitudes and masses and have noted, 
in agreement with the results of \citet{gag95a} and \citet{gar00}, a clear
dependence of the percentage of detections on stellar mass, from $\sim 0$\% at
the brown dwarf limit, to $\sim 100$\% for $\sim 2.0 M_\odot$ stars.  For 2 --
10 $M_\odot$ stars, detection fraction drops sharply, followed by a marked  increase (to
100\%) for the six highest mass ONC stars.

We concluded here by estimating activity indicators ($L_X$ and $L_X/L_{bol}$)
for stars in our {\em optical sample}; Paper~II studies correlations between
this X-ray activity and other stellar parameters. Our X-ray luminosities were
computed from {\em basal} count rates, assuming single temperature spectra for
all sources and proportionality between X-ray absorption and optical extinction,
with these assumptions verified by our analysis of medium resolution archival ACIS X-ray
spectra for a subset of our sources.

\section*{Acknowledgments}

The authors would like to thank M. Reid for providing his catalog of radio
sources in the ONC area prior to publication.

This work was partially supported at the CfA by NASA contracts NAS8-38248 and
NAS8-39073 and by NASA grant NAS5-4967. F.D., E.F., G.M., and S.S. wish to
acknowledge support from the Italian Space Agency (ASI) and MURST. E.F. would
like to thank the CfA for its hospitality during his Fellow visits.

\appendix

\section{The ACIS data \label{sect:ACIS_data}}

On 1999 October 12-13, the Advanced CCD Imaging Spectrometer (ACIS,
\citealt{tow00}) observed the central region of the ONC
for a total exposure time of $\sim 47$ ksec (``Obs ID'' 18, Sequence No. 200016). These
data were discussed by \citet{gar00} and are available in the {\em
Chandra} archive.

Here we utilize ACIS medium-resolution spectra
to demonstrate that the fluxes of
our HRC sources can be adequately described by an isothermal ($kT = 2.16$)
optically thin plasma, absorbed by a gas column density proportional to optical
extinction: $N_H= 2 \cdot 10^{21} A_V$.  This verifies the assumptions of \S
\ref{sect:Xprop} for a range of temperatures ($kT > 0.5$ keV) and for adherence
to the relationship between X-ray absorbing column and optical extinction.

Our spectral analysis characterized source spectra with two hardness ratios,
$HR_1$ and $HR_2$, and compared the measured $HR$ pairs with model predictions
for a grid of $kT$ and $N_H$ values. This approach is justified by the fact
that large statistical uncertainties for the majority of our
sample\footnote{For the brightest sources a more detailed spectral analysis can
be performed.} masks any (smaller) systematic inaccuracies introduced by our
simplifying assumptions.

\subsection{Data processing}

In order to improve upon the standard energy calibration, we applied the
{\sc correct\_cti} procedure \citep{tow00} to the Level 1 event file. This
procedure (partially) corrects for the spatial nonuniformity of the ACIS
spectral response that stems from degradation of the Charge Transfer
Inefficiency (CTI) during the early days of the {\em Chandra} mission. With
this correction, spectral resolution is $\sim$ 100-200~eV, 
depending both on the incidence position with respect to the
readout node and on the photon energy.
Following this correction we applied
standard grade and status flag filtering to produce a ``clean'' Level 2 event file.

We then extracted background-subtracted source counts in three spectral bands:
0.5-1.7keV (L), 1.7-2.8keV (M) and 2.8-8.0keV (H), for 678 HRC sources that
fell in the ACIS FOV. Source spectra were extracted from circles centered on
the HRC source positions, with radius 3 times the HRC source size as measured
by {\sc pwdetect} (see \S \ref{sect:detections}). Background was estimated from
concentric annuli of inner and outer radii of 4 and 5 times the HRC source
size. To avoid contamination effects, we discarded 36 sources because of the
proximity (distance $<$ 6 times the size) of another HRC source, and 150
additional sources were excluded due to net counts $\lesssim 0$ in one or more
of the three spectral bands.

For the remaining 492 sources we calculated two hardness ratios defined as
$HR_1=(M-L)/(M+L)$ and $HR_2=(H-M)/(H+M)$.  Uncertainties on the hardness
ratios were derived assuming that the number of source and background photons
in each spectral band followed a Poisson distribution with mean equal to the
measured value.

\subsection{Interpretation \label{sect:app_int}}

\begin{figure}
\centerline{\psfig{file=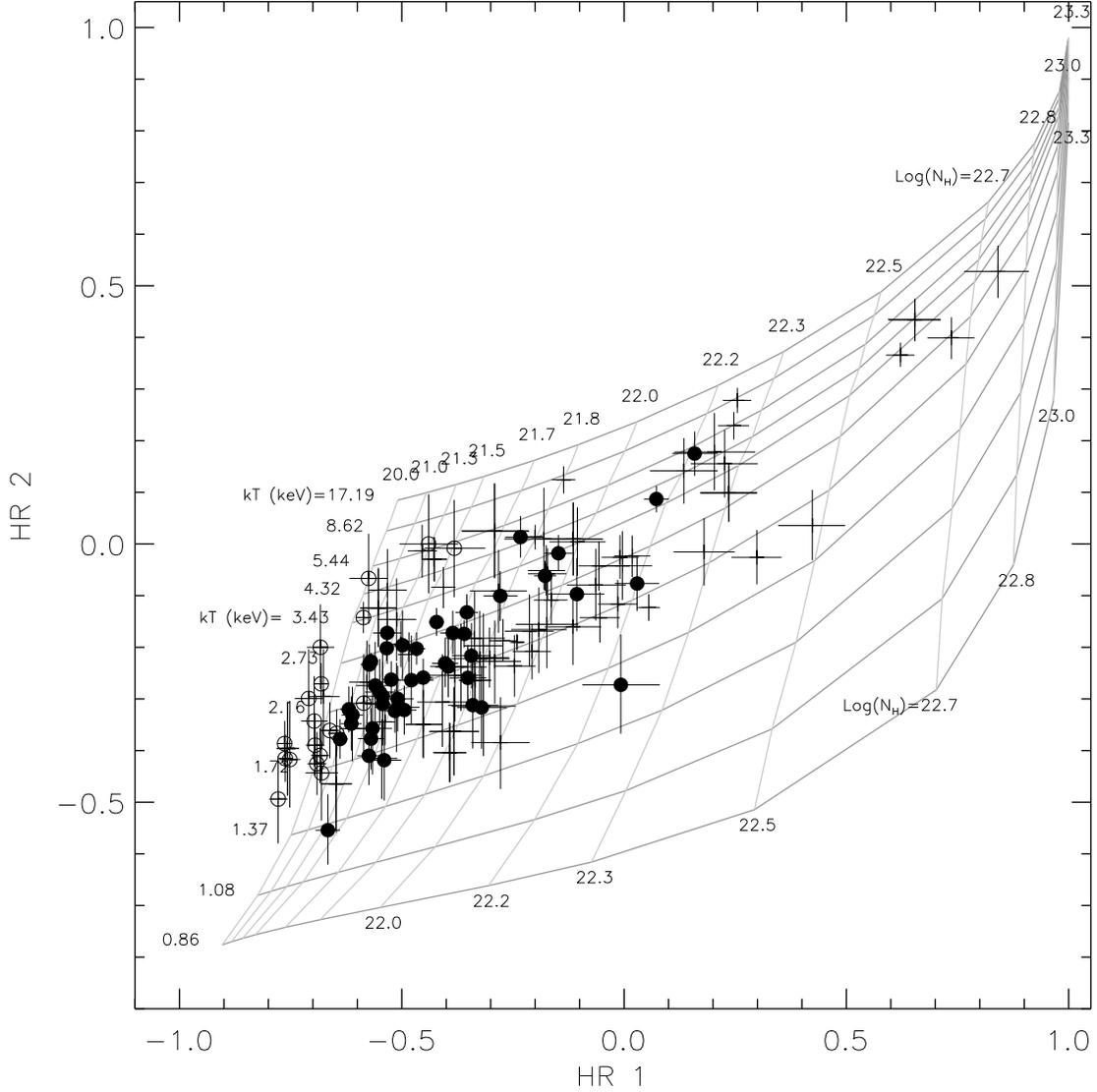,width=16cm}}
\caption{Grid of predicted ACIS hardness ratios for absorbed
single-temperature Raymond-Smith models; kT ranges from 0.27 to 8.62
keV, with $Log(N_H)$ running from 20 to 23.3.  Filled
circles denote {\em optical sample} stars that have hardness ratios
and errors within the grid, and whose inferred values of $kT$, $N_H$
and HRC conversion factor were used for our analysis (see text). Open circles have HR values (or errors)
near the grid edge, preventing their use.  Pluses denote positions (and
error bars) of HRC sources with computed errors of less than 0.1 for
$HR_1$ and $HR_2$.  \label{fig:HR_tk}}
\end{figure}

We computed hardness ratio predictions for a grid of isothermal Raymond-Smith
models with a range of temperatures and hydrogen column densities, using
PIMMS\footnote{Since PIMMS ignores the finite spectral resolution of ACIS, we
chose spectral bands to minimize the potential effect of significant spectral
features appearing in the selected passband boundaries. To confirm that we have
indeed avoided this problem we computed focal plane spectra for non-absorbed
sources for each grid temperature, convolved these spectra with Gaussians of
varying FWHM and then compared the hardness ratios obtained from these smoothed
spectra to those obtained from unsmoothed spectra. For a FWHM of 300 eV,
the difference between the two HR's is less than 0.023 in all cases,
negligible compared to typical uncertainties in our measured HR values.}.
Figure \ref{fig:HR_tk} shows the results of our PIMMS calculations in a grid of
$HR_1$ vs. $HR_2$; also shown are positions of high SNR HRC sources (see
caption). By interpolation in this plane, we can derive $kT^{HR}$ and
$N_H^{HR}$, where the superscript refers to the method used to estimate these
quantities. From these two spectral parameters we can derive the desired HRC
conversion factors, $CF^{HR}$ (cf. Figure \ref{fig:HRCcf}).

Formal $1\sigma$ errors on $kT^{HR}$, $N_H^{HR}$ and $CF^{HR}$ were derived
(along with uncertainties on the hardness ratios) by Monte Carlo methods with
the assumption that extracted source (and background) counts were Poisson
distributed with mean equal to the measured value. In order to compute mean
values and errors we require that the source has at least a 90\% chance of lying
inside the HR grid. For this reason a few points that lie inside the grid, but
close to its boundaries are excluded from the subsequent analysis, along with
those that lie outside. These latter points may be explained either by
statistical fluctuation in the hardness ratio or by the fact that their spectra
cannot be modeled as simply as we have assumed (e.g., they have a significant
second temperature thermal component; this could explain the points on the left
hand side of Fig. \ref{fig:HR_tk}). 

The majority of our sources have low SNR, in both the HRC and the ACIS
datasets.  This affects our hardness ratio analysis: the $1\sigma$ error bars
are in general quite large. Examination of Figure \ref{fig:HR_tk} shows that
(especially at very low and very high values) $kT^{HR}$ and $N_H^{HR}$ are quite
strong functions of the two ratios, making it difficult to determine the two
parameters. We therefore decided to include in the following conversion factor
analysis only sources with errors on $HR_1$ and $HR_2$ less than 0.1, i.e., the
subsample depicted in Figure \ref{fig:HR_tk}. This choice is dictated by our interest
in trends of median
$CF^{HR}$ with mass, accretion, etc.  By definition, the median is not weighted
and does not take into account variable errors, and including a larger number of
low SNR points would not necessarily improve the statistical significance of the median.

The median $kT^{HR}$ for our high SNR group is 2.2 keV with 90\% of the sources
in the 1.6-3.5 keV range. A fraction of this scatter can be explained by
uncertainties on the hardness ratios: with the mean temperature as reference
(i.e., $\sim 2.4$ keV), the formal $\chi^2$ is only 2.8. We examined trends in
$kT$ with stellar mass and age, finding no significant evidence for a
dependence in the range $0.2-10 M_{\odot}$. Neither could we find any
significant difference in the kT distributions of ``high'' and ``low''
accretion stars as defined from the Ca II equivalent width (cf. Paper II).
Given the range of temperatures found and the discussion concerning Figure
\ref{fig:HRCcf} (cf. \S \ref{sect:conv_fact}), we can be assured that an error
on the the assumed kT does not influence the HRC conversion factor by more than
$\sim 10$\%. Consequently, the most important aspect of our analysis becomes
the study of the relation between $N_H^{HR}$ and $A_V$.

The mean ratio between $A_V$ and $N_H^{HR}$ is $0.54
\cdot10^{-21}$, not very different from the nominal value $0.5\cdot10^{-21}$.
However there are several objects that lie far from this mean relation so that
the quartiles of the ratio distribution are $0.21\cdot10^{-21}$ and
$0.77\cdot10^{-21}$. It is unclear whether this large spread of values is due to
a real departure from the $N_H$ vs. $A_V$ mean relation (i.e., to an anomalous
gas to dust ratio) or to other factors such as an oversimplified spectral model.

\begin{figure}[!t!]
\centerline{\psfig{file=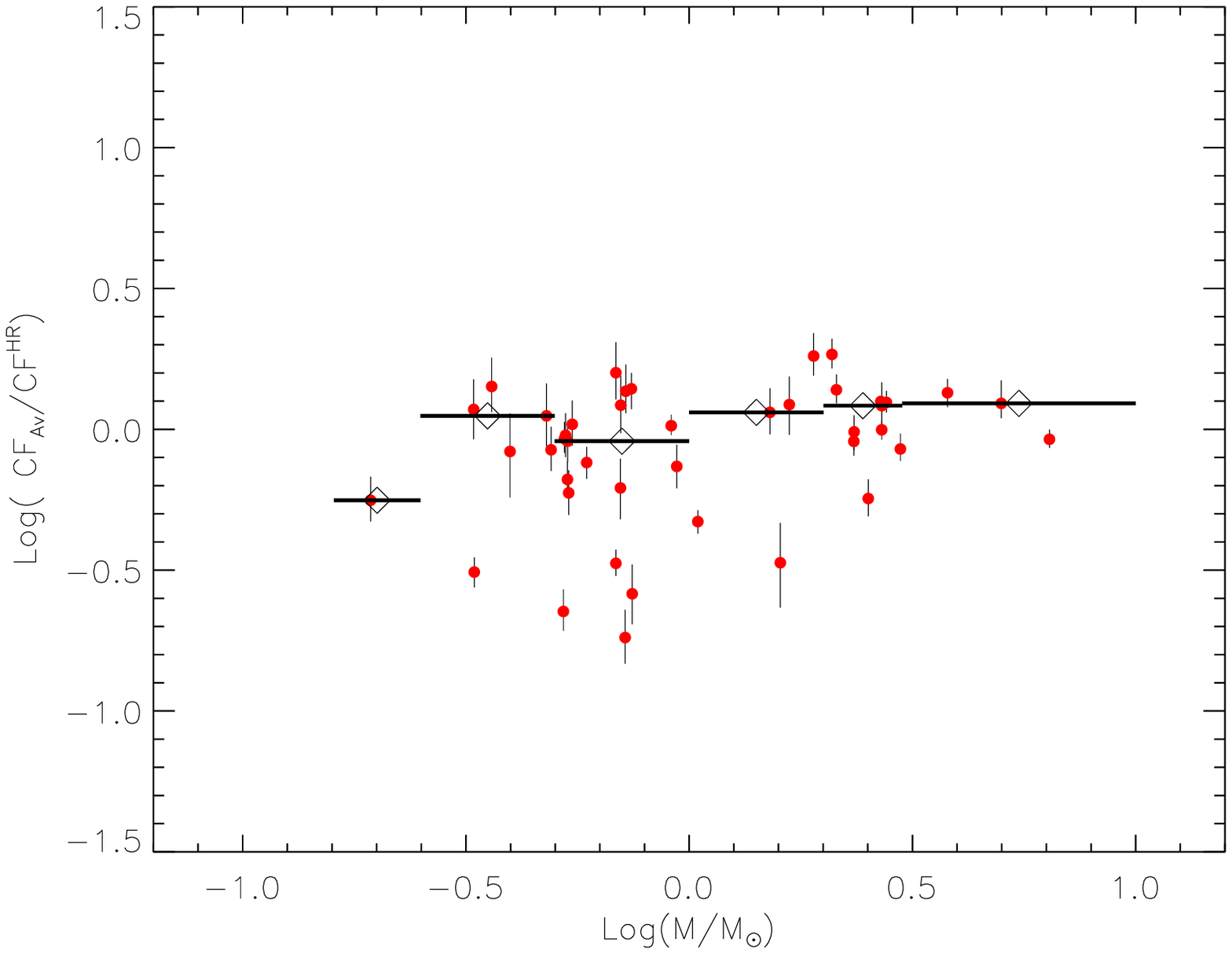,width=13cm}}
\caption{Ratio of HRC count rate to flux conversion factors vs. mass, computed
from optical extinction and $kT=2.16$keV as described in the text. Only stars
with high SNR are shown, with vertical error bars propagated from Poisson
statistics. Horizontal segments with diamonds in their center represent median
values of the points in the corresponding mass ranges (as employed in Paper~II). \label{fig:CF_ratio}}
\end{figure}

To conclude we examine the most important relation for our purpose: i.e., the
relation between conversion factors derived from hardness ratios, $CF^{HR}$,
and those derived assuming a constant temperature ($kT = 2.16$) and $N_H$
proportional to $A_V$ ($CF_{A_V}$). The median for the logarithm of the
$CF^{HR}/CF_{A_V}$ ratio is -0.024, and the $1\sigma$ quantiles are -0.252 and
0.130. Figure \ref{fig:CF_ratio} shows this ratio as a function of stellar mass
for our high SNR sample, as well as median values in the mass ranges defined in
Paper II and depicted in the figure as horizontal segments. No significant trend in the median is
apparent. We also considered separately samples of ``high accretion'' and ``low accretion''
stars, finding no difference in the conversion factor ratio;
sample sizes for determining this latter comparison, however, are very small: nine ``high accretion''
stars vs. 14 with ``low accretion'' (cf. Paper II).


\newpage 

\begin{thebibliography}{} 

\bibitem[Bally et al.(2000)]{bal00} Bally,  J., O'Dell, C.\ R., \& McCaughrean, M.\ J.\ 2000, \aj, 119, 2919 
\bibitem[Bessell(1991)]{bes91} Bessell, M.\ S.\ 1991, \aj,  101, 662 
\bibitem[Carpenter(2000)]{car00} Carpenter, J.\ M.\ 2000,  \aj, 120, 3139 
\bibitem[Carpenter et al.(2001)]{car01} Carpenter, J.~M., Hillenbrand, L.~A., \& Skrutskie, M.~F.\ 2001, \aj, 121, 3160
\bibitem[Damiani et al.(1997)]{dam97} Damiani, F., Maggio, A., Micela, G., \& Sciortino, S.\ 1997, \apj, 483, 350
\bibitem[Damiani et al.(2002)]{dam02} Damiani, F.\ et al. \ 2002 (In preparation) 
\bibitem[D'Antona \& Mazzitelli(1994)]{dan94} D'Antona, F.\  \& Mazzitelli, I.\ 1994, \apjs, 90, 467 
\bibitem[D'Antona \& Mazzitelli(1997)]{dan97} D'Antona F., Mazzitelli I.,  1997, Mem.\,S.\,A.\,It. Vol. 68. No. 4 (G. Micela, R. Pallavicini, S. Sciortino eds.)
\bibitem[Felli et al.(1993)]{fel93} Felli, M., Taylor, G.\  B., Catarzi, M., Churchwell, E., \& Kurtz, S.\ 1993, \aaps, 101, 127 
\bibitem[Flaccomio et al.(2002)]{fla02}  Flaccomio, E., Damiani, F., Micela, G., Sciortino, S., Harnden, F.~R., Murray, S. S., Wolk, S. J.\ 2002, ??
\bibitem[Gagn\'e et al.(1995)]{gag95a} Gagn\'e,  M., Caillault, J., \& Stauffer, J.\ R.\ 1995, \apj, 445, 280 
\bibitem[Garmire et al.(2000)]{gar00} Garmire, G., Feigelson,  E.\ D., Broos, P., Hillenbrand, L.\ A., Pravdo, S.\ H., Townsley, L., \&  Tsuboi, Y.\ 2000, \aj, 120, 1426 
\bibitem[Giacconi et al.(1972)]{gia72} Giacconi, R., Murray, 
S., Gursky, H., Kellogg, E., Schreier, E., \& Tananbaum, H.\ 1972, \apj, 
178, 281 
\bibitem[Goldsmith, Bergin, \& Lis(1997)]{gol97} Goldsmith,  P.~F., Bergin, E.~A., \& Lis, D.~C.\ 1997, \apj, 491, 615 
\bibitem[Harnden et al.(2001)]{frh01} Harnden, F.\ R.\ Adams, N. R., Damiani, F., Drake, J. J., Evans, N. R., Favata, F., Flaccomio, E., Freeman, P., Jeffries, R. ., Kashyap, V., Micela, G., Patten, B. M., Pizzolato, N., Schachter, J. F., Sciortino, S., Stauffer, J., Wolk, S. J., Zombeck, M. V. \ 2001, \apjl, 547, L141 
\bibitem[Herbst et al.(2000)]{her00} Herbst, W., Rhode, K.\ L., Hillenbrand, L.\ A., \& Curran, G.\ 2000, \aj,  119, 261 
\bibitem[Herbst et al.(2001)]{her01} Herbst, W., Bailer-Jones, C.~A.~L., \& Mundt, R.\ 2001, \apjl, 554, L197 
\bibitem[Hillenbrand(1997)]{hil97} Hillenbrand, L.\ A.\ 1997,  \aj, 113, 1733 
\bibitem[Hillenbrand \& Carpenter(2000)]{hil00} Hillenbrand,  L.\ A.\ \& Carpenter, J.\ M.\ 2000, \apj, 540, 236 
\bibitem[Hillenbrand et al.(1998)]{hil98a} Hillenbrand, L.\  A., Strom, S.\ E., Calvet, N., Merrill, K.\ M., Gatley, I., Makidon, R.\  B., Meyer, M.\ R., \& Skrutskie, M.\ F.\ 1998, \aj, 116, 1816 
\bibitem[Hillenbrand \& Hartmann(1998)]{hil98b} Hillenbrand,  L.\ A.\ \& Hartmann, L.\ W.\ 1998, \apj, 492, 540 
\bibitem[Houdashelt et al.(2000)]{hou00}  Houdashelt, M.\ L., Bell, R.\ A., \& Sweigart, A.\ V.\ 2000, \aj, 119, 1448
\bibitem[Jones \& Walker(1988)]{jon88} Jones, B.\ F.\ \&  Walker, M.\ F.\ 1988, \aj, 95, 1755 
\bibitem[Ku \& Chanan(1979)]{ku79} Ku, W.~H.-M.~\& Chanan, G.~A.\ 1979, \apjl, 234, L59 
\bibitem[Lada et al.(2000)]{lad00} Lada, C.\ J., Muench, A.\  A., Haisch, K.\ E., Lada, E.\ A., Alves, J.\ F., Tollestrup, E.\ V., \&  Willner, S.\ P.\ 2000, \aj, 120, 3162 
\bibitem[Lucas et al.(2001)]{luc01} Lucas, P.~W., Roche, P.~F., Allard, F., \& Hauschildt, P.~H.\ 2001, \mnras, 326, 695
\bibitem[Lucas \& Roche(2000)]{luc00} Lucas, P.\ W.\ \&  Roche, P.\ F.\ 2000, \mnras, 314, 858 
\bibitem[Luhman et al.(2000)]{luh00} Luhman, K.~L., Rieke,  G.~H., Young, E.~T., Cotera, A.~S., Chen, H., Rieke, M.~J., Schneider, G.,  \& Thompson, R.~I.\ 2000, \apj, 540, 1016 
\bibitem[Murray et al.(2000)]{mur00} Murray, S.\ S.\ et al.\  2000, \procspie, 4012, 68 
\bibitem[O'Dell(2001)]{ode01} O'Dell, C.\ R.\ 2001, \pasp,  113, 29 
\bibitem[Rebull(2001)]{reb01} Rebull, L.\ M.\ 2001, \aj, 121,  1676 
\bibitem[Ryter (1996)]{ryt96} Ryter, CH.E., 1996, Ap\&SS 236, 285
\bibitem[Schulz et al.(2001)]{sch01} Schulz, N.\ S.,  Canizares, C., Huenemoerder, D., Kastner, J.\ H., Taylor, S.\ C., \&  Bergstrom, E.\ J.\ 2001, \apj, 549, 441 
\bibitem[Siess, Dufour, \& Forestini(2000)]{sie00} Siess, L.,  Dufour, E., \& Forestini, M.\ 2000, \aap, 358, 593 
\bibitem[Simon, Dutrey, \& Guilloteau(2000)]{sim00} Simon,  M., Dutrey, A., \& Guilloteau, S.\ 2000, \apj, 545, 1034 
\bibitem[Stassun et al.(1999)]{sta99}  Stassun, K.\ G., Mathieu, R.\ D., Mazeh, T., \& Vrba, F.\ J.\ 1999, \aj,  117, 2941 
\bibitem[Townsley et al.(2000)]{tow00}  Townsley, L.~K., Broos, P.~S., Garmire, G.~P., \& Nousek, J.~A.\ 2000,  \apjl, 534, L139 
\bibitem[Wainscoat et al.(1992)]{wai92} Wainscoat, R.\ J.,  Cohen, M., Volk, K., Walker, H.\ J., \& Schwartz, D.\ E.\ 1992, \apjs, 83,  111 
\bibitem[Weisskopf et al.(2002)]{wei02} Weisskopf, M.~C., Brinkman, B., Canizares, C., Garmire, G., Murray, S.~S. and Van Speybroeck, L.~P. 2002, PASP, 114, 1

\end{thebibliography}
\end{document}